\documentclass[12pt]{article}
\usepackage{graphicx}
\usepackage{amsmath}
\usepackage{amssymb}
\usepackage{ulem}
\usepackage{enumerate}
\usepackage{cite}
\usepackage{latexsym}
\begin{document}
\begin{center}
{\bf {\large{Canonical Brackets, (Anti-)BRST Invariant Charges and Physicality Criteria: Non-Abelian 1-Form Gauge Theory }}}

\vskip 3cm

{\sf  R. P. Malik$^{(a,b)}$}\\
$^{(a)}$ {\it Physics Department, Institute of Science,}\\
{\it Banaras Hindu University, Varanasi - 221 005, India}\\

\vskip 0.1cm

$^{(b)}$ {\it DST Centre for Interdisciplinary Mathematical Sciences,}\\
{\it Institute of Science, Banaras Hindu University, Varanasi - 221 005, India}\\
{\small {\sf {e-mails: rpmalik1995@gmail.com; malik@bhu.ac.in}}}
\end{center}

\vskip 1.5 cm

\noindent
{\bf Abstract:}
In the case of a D-dimensional non-Abelian 1-form gauge theory (without any interaction with the matter fields), 
we show that the application of the Noether theorem does {\it not}
lead to the derivations of the Becchi-Rouet-Stora-Tyutin (BRST) and anti-BRST charges that obey (i) the (anti-)BRST invariance, and
(ii) the nilpotency property (unless we exploit the theoretical strength of $(a)$ the 
partial integration along with the Gauss divergence theorem, and $(b)$ use 
the appropriate equations of motion at suitable places). 
This happens because of the presence of a {\it non-trivial} Curci-Ferrari (CF) condition on our BRST-quantized D-dimensional
non-Abelian 1-form gauge theory (whose limiting case is the BRST-quantized D-dimensional Abelian 
1-form gauge theory where the CF-type restriction  is {\it trivial} and the 
corresponding Noether (anti-)BRST charges
turn out to be nilpotent as well as (anti-)BRST invariant quantities {\it together}). We exploit the
theoretical strength of the {\it basic} canonical (anti)commutators to prove (i) the Noether (anti-)BRST charges as the true generators for
the off-shell nilpotent (anti-)BRST symmetry transformations, and
(ii) the (anti-)BRST invariance of the consistently {\it modified} versions of the Noether (anti-)BRST charges. We 
demonstrate that the (anti-)BRST invariant versions of the conserved (anti-)BRST charges are useful in the discussion on the physicality criteria
because of their consistency with the Dirac quantization conditions for the gauge theories.



\newpage
\section {Introduction}

The principle of local gauge invariance is at the heart of the {\it interacting} (non-)Abelian 1-form gauge theories that provide a precise 
theoretical description\footnote{One of the cardinal examples for such kinds of gauge theories is the well-known 4D standard model of elementary particle physics
where the agreements between theory and experiment are outlandish in nature. This theory, despite being plagued by a few shortcomings, 
provides the cornerstone for the {\it unified} theoretical description of the electromagnetic, weak and strong fundamental interactions of nature.}
of the {\it three} fundamental interactions of nature. One of the mathematically  elegant and 
physically intuitive methods
to covariantly quantize such kinds of gauge theories is the Becchi-Rouet-Stora-Tyutin (BRST) formalism [1-4] where the unitarity and quantum gauge 
(i.e. BRST) invariance are respected  {\it together} (see, e.g. [5] for more details) at any arbitrary order of perturbative computations 
for a given physical process (that is allowed by the relevant  {\it interacting} gauge theory). The validity of unitarity is {\it primarily} ensured by the 
presence of the (anti-)ghost fields in a given  BRST-quantized D-dimensional interacting gauge theory [5].

The {\it classical} interacting gauge theories always respect the infinitesimal, continuous and {\it local} gauge symmetry transformations 
that are generated by the first-class constraints (in the terminology of
Dirac's prescription for the classification scheme) which exist on {\it these}
 theories. Within the framework of the BRST formalism, the {\it local} gauge symmetry transformations
are traded with the infinitesimal, continuous and nilpotent BRST and anti-BRST symmetry transformations. These nilpotent symmetry transformation operators
are required to absolutely anticommute with each-other so that they can maintain their own independent identity. The nilpotency and absolute
anticommutativity properties  are a couple of sacrosanct requirements of the quantum gauge (i.e. BRST and anti-BRST) symmetry  transformation operators.
In particular, the requirement of the absolute anticommutativity property of the (anti-)BRST transformation operators {\it distinguishes} them
from the $\mathcal{N} = 2$ SUSY transformation operators which are {\it also} nilpotent of order two. However, these set of 
$\mathcal{N} = 2 $ SUSY transformation
operators do {\it not} anticommute with each-other. This {\it specific} feature is {\it one} of the sacrosanct properties of the $\mathcal{N} = 2 $
SUSY transformation operators.

The key motivating factors behind our present investigation are as follows. In our recent earlier work [6], we have established that
the {\it Noether} conserved (anti-)BRST charges are {\it always} found to be non-nilpotent\footnote{This statement is true when we take into account
the well-known relationship between the infinitesimal, continuous and off-shell nilpotent (anti-)BRST symmetry transformation operators and their
generators as the conserved Noether (anti-)BRST charges  in their operator forms [cf. Eq. (12) below]. }
if the BRST-quantized theory is endowed with the
{\it non-trivial} CF-type restriction(s). Furthermore, {\it these} Noether charges are {\it also} found to be non-invariant under the 
{\it off-shell} nilpotent (anti-)BRST
transformations  [cf. Eq. (12) below]. As a consequence, these Noether charges are {\it not} physical (in the true sense of the word within the 
framework of BRST formalism).
These non-nilpotent (anti-)BRST charges, however, are {\it useful} in the sense that they are
the generators for the infinitesimal, continuous and {\it off-shell} 
nilpotent (anti-)BRST symmetry transformations (from which they are derived). It is 
crystal clear that such kinds of conserved Noether
(anti-)BRST charges (which are found to be non-invariant under the {\it off-shell} nilpotent (anti-)BRST symmetry transformations)
can {\it not} be used in the discussion on the physicality criteria 
within the framework of BRST formalism (see, e.g. [5,7]). Hence, one has to
use the theoretical strength of (i) the partial integration along with the Gauss divergence theorem, and (ii) 
appropriate Euler-Lagrange (EL) equations of motion (EoM) at suitable places, 
to recast these {\it non-invariant}  charges in the forms that are found to be
(anti-)BRST invariant. In our earlier works [8-10], we have been able to establish that the (anti-)BRST 
{\it invariant} versions of the (anti-)BRST charges are useful
in the physicality criteria because they lead to the derivation of the  conditions on the physical states that are found to be 
consistent with the Dirac quantization conditions [11,12] for the gauge theories (which are endowed with the first-class constraints 
in the terminology of Dirac's prescription for the classification scheme). In other words, we find that the physicality criteria w.r.t. 
the (anti-)BRST {\it invariant} versions of the conserved (anti-)BRST charges lead to
the annihilation of the physical states by the operator forms of the first-class constraints (cf. Appendix A below) of the original {\it classical}
D-dimensional non-Abelian 1-form gauge theory at the {\it quantum} level.

The central purpose of our
present investigation  is to exploit the theoretical strength of the {\it basic} canonical (anti)commutators to prove (i) the conserved
Noether (anti-)BRST charges to be the true generators for the infinitesimal, continuous and off-shell nilpotent 
(anti-)BRST symmetry transformations, and (ii) the (anti-)BRST invariance\footnote{ One of the highlights of our present endeavor
is the proof of the (anti-)BRST invariance of the consistently {\it modified} versions of the conserved (anti-)BRST charges by using the 
basic canonical (anti)commutators of our present BRST-quantized non-Abelian 1-form theory (cf. Subsec. 5.2 below).}  
of the consistently {\it modified} versions of the Noether
(anti-)BRST charges. We demonstrate that, in the physicality criteria (cf. Subsec. 5.3 below for details), the (anti-)BRST invariant
versions of the (anti-)BRST charges play a decisive role because they lead to the conditions on the true {\it physical} states
(existing in the {\it total} quantum Hilbert space of states) that are found to be consistent with the Dirac quantization conditions
on our BRST-quantized D-dimensional non-Abelian 1-form gauge theory [which is endowed with the first-class constraints 
(cf. Appendix A below) at the {\it classical} level]. To be precise, we observe that the operator forms of the first-class
constraints annihilate the {\it true} physical states at the quantum level which are nothing but the Dirac quantization conditions
(see, e.g. [12] for details). On the contrary, when we invoke the Noether conserved (anti-)BRST charges in the physicality criteria, 
we end up with a set of quite {\it absurd}  results (cf. Subsec. 5.3 below for details).

The theoretical contents of our present endeavor are organized as follows. In the next section, we briefly recapitulate the bare essentials of the 
nilpotent (anti-)BRST symmetry transformations and derive the Noether conserved (anti-)BRST charges. The subject matter of our Sec. 3 is related 
to the canonical method of the proof that the Noether conserved (anti-)BRST charges are the true generators for the 
off-shell nilpotent (anti-)BRST symmetry transformations.
Our Sec. 4 is devoted to the proof that
{\it these} charges are {\it not} nilpotent as well as (anti-)BRST invariant  when we use the off-shell nilpotent symmetry transformations and their 
generators as the Noether conserved (anti-)BRST charges. We exploit (i) the partial integration along with  the Gauss divergence theorem, and (ii)
the appropriate EL-EoM at suitable places,  
to prove the (anti-)BRST invariance as well as the nilpotency property of the conserved Noether (anti-)BRST charges. Sec. 5 of our
present endeavor deals with the proof that the consistently {\it modified} versions of the Noether conserved 
(anti-)BRST charges (i) satisfy the property of (anti-)BRST invariance, and (ii) participate {\it fruitfully} in the 
physicality criteria. Finally, in Sec. 6, we summarize our key observations
and point out the future perspective of our present investigation.

In our Appendix A, we discuss the first-class constraints of our {\it classical} D-dimensional non-Abelian 1-form gauge theory
(without any interaction with {\it matter} fields).
These constraints are useful in (i) the main body of our text,  and (ii) our key discussions related to the physicality criteria. 
Our Appendix B is devoted to the derivations of the CF-condition [15] from a couple of different theoretical angles. In Appendix C, 
we collect a few standard rules of the (anti)commutators that have been used in the main body of the text. Our Appendix D deals with
a few explicit computations that establish that the Noether conserved (anti-)BRST charges are the generators for the off-shell nilpotent 
(anti-)BRST symmetry transformations of the basic {\it dynamical} fields of our BRST-quantized theory.\\

{\it Conventions and Notations:} In the entire text, we adopt the convention of the left derivative w.r.t. {\it all} the fermionic fields of our theory
in {\it all} appropriate computations (e.g. the canonical conjugate momenta, 
EL-EoM, Noether theorem, etc.). Our D-dimensional flat Minkowskian spacetime manifold is endowed with the
metric tensor $\eta_{\mu\nu} = $ diag (+1, -1, -1...) where the Greek indices $\mu, \nu, \lambda...= 0, 1, 2...(D-1) $ stand for the time
and space directions. We choose the Latin indices $i, j, k...= 1, 2...(D-1)$ to denote the space directions {\it only}. While applying
the Gauss divergence theorem, we assume that there is {\it no} non-trivial topology at the boundary. The 
infinitesimal, continuous and nilpotent
(anti-)BRST transformation operators are represented by the notations $s_{(a)b} $ and the corresponding Noether conserved (anti-)BRST charges
are denoted by the symbols $Q_{(a)b}$. The over dot (i.e. $\dot \Phi $) on  the generic field $\Phi$ 
 of our theory stands for the time derivative
(i.e. $ \partial_0 \Phi  = \partial \Phi/\partial t$) in the natural units $\hbar = c = 1 $. We freely use the brackets $(...)$ and  $[...] $
in the text for convenience. However, at appropriate places in the text and equations, the brackets $\big(\{...\}\big) [...] $ {\it also}
stand for the (anti)commutators depending on the context.\\


\section{Preliminary: Noether (Anti-)BRST Charges }

We focus on  the 
coupled (but equivalent)  Lagrangian densities for a D-dimensional
non-Abelian 1-form gauge theory (see, e.g. [5]) in the Curci-Ferrari (CF) gauge [13,14] where there is {\it no}
 interaction between the gauge field and any kinds of matter fields.
The  explicit expressions for these properly {\it gauge-fixed} Lagrangian densities (that incorporate into themselves the appropriate forms of the 
Faddeev-Popov (FP) ghost terms) are
\begin{eqnarray}\label{1}
{\cal L}_{(B)} = -\, \frac{1}{4}\, F^{\mu\nu} \cdot F_{\mu\nu} + B\, \cdot (\partial_\mu\, A^\mu) 
+ \frac{1}{2}\, (B\cdot B + \bar B\cdot \bar B) -\, i\, \partial_\mu\bar C \cdot D^\mu\, C, \nonumber\\
{\cal L}_{(\bar B)} = -\, \frac{1}{4}\, F^{\mu\nu} \cdot F_{\mu\nu} - \bar B\, \cdot (\partial_\mu\, A^\mu) 
+ \frac{1}{2}\, (B\cdot B + \bar B\cdot \bar B) -\, i\, D_\mu\, \bar C \cdot \partial^\mu C,
\end{eqnarray}
where the field strength tensor $F_{\mu\nu}^a = \partial_\mu\, A_\nu^a - \partial_\nu\, A_\mu^a + i\, f^{abc}\, A_\mu^b\, A_\nu^c$
($\mu, \nu...= 0, 1, 2...D-1 $)
has been derived from the non-Abelian 2-form: $F^{(2)} = d\,A^{(1)} + i\, A^{(1)} \wedge A^{(1)}$ where the  1-form 
$A^{(1)} = \, A_\mu \,(d x^\mu) \equiv   (A_\mu^a\, T^a)\, (d\, x^\mu)$ defines the non-Abelian gauge field ($A_\mu^a$)
and $d = \partial_\mu \, (dx^\mu)$ [with $d^2 = \frac{1}{2!}\,(\partial_\mu \partial_\nu - \partial_\nu \partial_\mu)\, (d x^\mu \wedge d x^\nu) = 0 $]
is the exterior derivative of differential geometry.
For the $SU(N)$ Lie algebraic space,
 we have the Lie algebra: $[T^a, \, T^b] = f^{abc}\, T^c$ that is satisfied by the $SU(N)$ generators $T^a\,(a, b, c... = 1, 2..., N^2-1)$.
Here $f^{abc}$ are the structure constants\footnote{ We adopt the dot and cross products {\it only} in the $SU(N)$ Lie algebraic space where 
 we have: $P\cdot Q = P^a\, Q^a, \; (P \times Q) = (P \times Q)^a\, T^a \equiv f^{abc}\, P^b\, Q^c\, T^a$ for the two non-null vectors $P^a$ 
and $Q^a$ in the $SU(N)$ Lie-algebraic space.
We take into account the summation convention where the repeated indices are summed over. It is 
crystal clear that (i) if the vectors $P^a$ and $Q^a$ are bosonic,
 we have: $(P\cdot Q) = (Q\cdot P) $ and $(P \times Q) = -\, (Q\times P)$, and (ii) if the vectors $P^a$ and $Q^a$ are fermionic,
 we have: $(P\cdot Q) = -\,(Q\cdot P) $ and $(P \times Q) = +\, (Q\times P)$. It is worthwhile to lay emphasis 
on the fact that the rules of the dot and cross products,
 for {\it both} the vectors being {\it bosonic}, are also {\it true} if one of them is fermionic and the other is bosonic. The covariant derivatives
  $D_\mu \bar C  = \partial_\mu \bar C + i\, (A_\mu \times \bar C)$ and
$D_\mu C  = \partial_\mu  C + i\, (A_\mu \times C)$ on the (anti-)ghost fields $(\bar C^a)C^a$
 are  in the adjoint representation of the  $SU(N)$ Lie algebra. } which can be chosen [7] to be {\it totally} antisymmetric in all the indices
 for the semi-simple Lie group $SU(N)$. In the above coupled Lagrangian densities (1), we have the Nakanishi-Lautrup auxiliary fields $B^a$
 and $\bar B^a$ that are restricted to satisfy the well-known CF-condition: $B + \bar B + (\bar C \times C) = 0$ [15] where the
 fermionic [i.e. $C^a \,C^b + C^b\, C^a = 0, \; \bar C^a\, \bar C^b + \bar C^b\, \bar C^a = 0, \; 
C^a \,\bar C^b + \bar C^b\, C^a = 0, \; (C^a)^2 = 0, (\bar C^a)^2 = 0 $, etc.] (anti-)ghost fields are denoted by $(\bar C^a)C^a$.
The action integrals corresponding to the   
coupled (but equivalent) Lagrangian densities (1) respect the following infinitesimal, continuous and off-shell
 nilpotent [$s_{(a)b} ^2 = 0$] (anti-)BRST {\it symmetry} transformations [$s_{(a)b}$], namely;
\begin{eqnarray}\label{2}
&&s_{ab}\; A_\mu= D_\mu\bar C,\qquad  s_{ab}\; \bar C= -\frac{i}{2}\,(\bar C\times\bar C),
\qquad s_{ab}\;C   = i{\bar B},\qquad  s_{ab}\;\bar B    = 0,\nonumber\\
&&s_{ab}\; F_{\mu\nu}      = i \,(F_{\mu\nu}\times\bar C),\qquad s_{ab} (\partial_\mu A^\mu) = \partial_\mu D^\mu\bar C,
\qquad s_{ab}\; B = i\,(B \times \bar C), \nonumber\\
&&s_b\; A_\mu = D_\mu C,\qquad s_b \;C =  - \frac{i}{2}\; (C\times C),\qquad  s_b\;\bar C \;= i\,B ,\;
 \qquad s_b\; B = 0, \nonumber\\
&& s_b\;\bar B = i\,(\bar B\times C),\qquad s_b\; (\partial_\mu A^\mu) = \partial_\mu D^\mu C,\qquad s_b \;F_{\mu\nu} = i\,(F_{\mu\nu}\times C),
\end{eqnarray}
because of the following observations: 
\begin{eqnarray}\label{3}
s_b {\cal L}_{(B)}  =  \partial_\mu(B \cdot D^\mu C),  \;\qquad\; s_{ab}{\cal L}_{(\bar B)}= - \;\partial_\mu{(\bar B \cdot D^\mu \bar C)}.
\end{eqnarray}
In other words, the action integrals (corresponding to the Lagrangian densities ${\cal L}_{(B)} $ and ${\cal L}_{(\bar B)} $) are {\it perfectly} 
BRST and anti-BRST invariant, respectively, 
 because we do {\it not} invoke any kinds of outside restrictions (e.g. the EL-EoM, CF-condition, etc.) for the  proof of their invariance
due to our observations in equation (3). The Gauss divergence theorem is, of course, taken into account for the proof of the {\it above} invariance.

According to Noether's theorem, the invariance (i.e. $s_b S_1 = 0 $ and $s_{ab} S_2 = 0 $) of the action integrals 
(i.e. $S_1 = \int d^D x\, {\cal L}_{(B)}  $ and $S_2 = \int d^D x\, {\cal L}_{(\bar B)} $) under the {\it perfect} BRST and anti-BRST symmetry transformations [cf. Eq. (3)] leads to the derivations of the
conserved  $\partial_\mu\, J_{(r)}^{\mu} = 0$ (with $r = b, \; ab$) Noether currents $J^\mu_{(b)}$ and $J^\mu_{(ab)}$ as follows:
\begin{eqnarray}\label{4}
&& J_{(b)}^\mu = B \cdot D^\mu\,  C - \, F^{\mu\nu} \cdot D_\nu\,  C
+ \frac{1}{2}\, \partial^\mu\, \bar C \cdot ( C \times C), \nonumber\\
&&J_{(ab)}^\mu = -\, F^{\mu\nu}\cdot D_\nu\, \bar C -\, \bar B \cdot D^\mu\, \bar C 
-\, \frac{1}{2}\, (\bar C \times \bar C)\cdot \partial^\mu\, C.
\end{eqnarray}
Following the sacrosanct prescription of Noether's theorem, we derive the expressions for the {\it conserved} (anti-)BRST 
charges ($Q_{(a)b} = \int d^{D-1} x\, J_{(a)b}^0)$ as follows
\begin{eqnarray}\label{5}
&&Q_{ab} = -\, \int d^{D-1} x\, \Big[ F^{0i}\cdot D_i\, \bar C + \bar B \cdot D_0\, \bar C 
+\, \frac{1}{2}\, \dot C \cdot (\bar C \times \bar C)    \Big], \nonumber\\
&&Q_{b} =  \int d^{D-1} x\, \Big[  B \cdot D_0\,  C - \, F^{0i} \cdot D_i\,  C
+ \frac{1}{2}\,  \dot {\bar C } \cdot ( C \times C)    \Big],
\end{eqnarray}
where $\dot C $ and $\dot {\bar C } $ denote the time derivatives (i.e. $\dot C = \partial_0 C $ and $\dot {\bar C } = \partial_0 \bar C $) on these fields. 
In the next section, we shall be able to show that the Noether conserved (anti-)BRST charges are the {\it true} generators for the off-shell nilpotent
(anti-)BRST symmetry transformations (2) where the {\it basic} canonical (anti)commutators will be invoked for the proof.

We end this short (but very useful) section with the following remarks. 
First, in the derivation of the EL-EoM and Noether conserved currents [cf. Eq. (4)], we have
taken into account the convention of the left derivative w.r.t. the fermionic fields. Second, for
 the proof of the conservation law $\partial_\mu\, J_{(r)}^{\mu} = 0$ (with $r = b, \; ab$) 
Noether currents $J^\mu_{(b)}$ and $J^\mu_{(ab)}$ [cf. Eq. (4)],  
we have used  the EL-EoM that have been derived from the coupled (but equivalent) Lagrangian  densities of (1). Third, the
Abelian limit of the CF-condition (i.e.  $B + \bar B + (\bar C \times C) = 0$) implies that we have a {\it single} Lagrangian density
because we obtain: ${\cal L}_{(B)} = {\cal L}_{(\bar B)} $. This is due to the fact that the SU(N) group of the non-Abelian theory reduces to the U(1)
gauge group where there are {\it no} group indices and, therefore, we end up with $\bar B = -\, B$. Fourth, the kinetic term
 (i.e. $-\, \frac{1}{4}\, F^{\mu\nu} \cdot F_{\mu\nu} $), owing its origin to the exterior derivative of differential geometry (see, e.g. [16-19]),
remains invariant under the off-shell nilpotent (anti-)BRST symmetry transformations. Finally, the Lagrangian densities in equation (1)
are {\it coupled} and {\it equivalent} because of the presence of the CF-condition [15] on our BRST-quantized 
non-Abelian 1-form gauge theory (cf. Appendix B for details).\\


\section{Basic Brackets: Covariant Canonical Quantization and Proof of the Noether Charges as Generators}

The purpose of this section is (i) to perform the covariant canonical quantization of the (anti-)BRST invariant [cf. Eq. (3)]
Lagrangian densities ${\cal L}_{(\bar B)}$ and ${\cal L}_{(B)} $ by taking into account the {\it basic} (anti)commutators of our
BRST-quantized non-Abelian 1-form gauge theory, and (ii) to demonstrate the usefulness of {\it these} basic brackets in the proof that
the Noether conserved (anti-)BRST charges $Q_{(a)b} $ [cf. Eq. (5)] are the generators for the 
infinitesimal, continuous and off-shell nilpotent (anti-)BRST transformations (2).
Toward this goal in mind, first of all, we define the canonical conjugate momenta (i.e. $\Pi $'s) w.r.t. the dynamical fields $A_\mu, C, \bar C$ of the
Lagrangian  density  ${\cal L}_{(B)}$ as
\begin{eqnarray}\label{6}
&& \Pi_{(A)}^{\mu a} (B) = \dfrac{\partial \, {\cal L}_{(B)}}{\partial\, (\partial_0\, A^a_\mu)} = -\, F^{0\mu a} + \eta^{0\mu} \, B^a
\;\;\Longrightarrow\;\; \Pi_{(A)}^{0 a} (B) = B^a, \qquad  \Pi_{(A)}^{i a} (B) = -\, F^{0i a}, \nonumber\\
&& \Pi^a_{(C)} (B) = \dfrac{\partial \, {\cal L}_{(B)}}{\partial\, (\partial_0\, C^a)} = i \, \partial_0 \bar C \equiv i\, \dot {\bar C}, \qquad
\Pi^a_{(\bar C)} (B) = \dfrac{\partial \, {\cal L}_{(B)}}{\partial\, (\partial_0\, \bar C)^a} = -\,i\, (D_0  C)^a,
\end{eqnarray} 
where (i) the subscripts on the conjugate momenta ($\Pi$'s) denote the shorthand notations for the {\it basic} dynamical fields $A_\mu, C, \bar C$
of our (anti-)BRST invariant theory, (ii) the canonical conjugate momenta, w.r.t. the fermionic (anti-)ghost fields, have been computed by
taking into account the convention of the left derivative, (iii) the
notation $\eta^{0\mu} $ is a component of the inverse (i.e. $\eta^{\mu\nu} $) of the flat metric tensor $\eta_{\mu\nu}$ = diag $(+1, -1, -1,......)$ 
that has been chosen for the
D-dimensional background Minkowskian spacetime manifold, and (iv) the parenthesis $(B)$ in front of the momenta ($\Pi$'s) denotes that these
momenta have been derived from the {\it perfectly} BRST invariant  [cf. Eq. (3)] 
Lagrangian density ${\cal L}_{(B)}$. In exactly similar fashion, we can define the conjugate momenta ($\Pi$'s), 
corresponding to the {\it basic} fields of the {\it perfectly} anti-BRST invariant [cf. Eq. (3)] 
Lagrangian  density  ${\cal L}_{(\bar B)}$, as
\begin{eqnarray}\label{7}
&& \Pi_{(A)}^{\mu a} (\bar B) = \dfrac{\partial \, {\cal L}_{(\bar B)}}{\partial\, (\partial_0\, A^a_\mu)} = -\, F^{0\mu a} - \eta^{0\mu} \, \bar B^a
\;\;\Longrightarrow\;\; \Pi_{(A)}^{0 a} (\bar B) = -\, \bar B^a, \qquad  \Pi_{(A)}^{i a} (\bar B) = -\, F^{0i a}, \nonumber\\
&& \Pi^a_{(C)} (\bar B) = \dfrac{\partial \, {\cal L}_{(\bar B)}}{\partial\, (\partial_0\, C^a)} = i \, (D_0 \bar C)^a, \qquad
\Pi^a_{(\bar C)} (\bar B) = \dfrac{\partial \, {\cal L}_{(\bar B)}}{\partial\, (\partial_0\, \bar C^a)} = -\,i\, \partial_0 C^a \equiv - \, i\, \dot C^a,
\end{eqnarray} 
where the symbols are clear (and we need not explain them once again). With the help of the conjugate momenta, defined
in equations (6) and (7), we can define the {\it basic} canonical (anti)commutators for the 
Lagrangian densities ${\cal L}_{(B)} $ and ${\cal L}_{(\bar B)}$, respectively.

In the natural units: $\hbar = 1, \; c = 1$, we have the  
{\it general} forms of the  {\it non-zero} equal-time basic canonical (anti)commutators, corresponding to the coupled Lagrangian densities ${\cal L}_{(B)} $
and ${\cal L}_{(\bar B)} $ for our D-dimensional BRST-quantized  non-Abelian 1-form gauge theory, as: 
\begin{eqnarray}\label{8}
&&\big \{ C^a (\vec{x}, \, t), \;\;\; \Pi^b_{(C)} (\vec{y},  \, t) \big \} = i \, \delta^{ab}\, \delta^{(D-1)} (\vec{x} - \vec{y}), 
\nonumber\\
&&\big \{ \bar C^a (\vec{x}, \, t), \;\;\; \Pi^b_{(\bar C)} (\vec{y},  \, t) \big \} = i \, \delta^{ab}\, \delta^{(D-1)} (\vec{x} - \vec{y}), 
\nonumber\\
&&\big [ A_{0}^a (\vec{x}, \, t), \;\;\;  \Pi^{0b}_{(A)} (\vec{y},  \, t) \big ] = i \, \delta^{ab}\, \delta^{(D-1)} (\vec{x} - \vec{y}), 
\nonumber\\
&&\big [ A_{i}^a (\vec{x}, \, t), \;\;\;\Pi^{jb}_{(A)} (\vec{y},  \, t) \big ] = i \, \delta^{ab}\,\delta_i^{j}\, \delta^{(D-1)} (\vec{x} - \vec{y}).
\end{eqnarray}
The {\it rest} of {\it all} the possible brackets (i.e. anticommutators and/or commutators) are {\it zero}. Written in terms of the {\it actual} expressions for the
momenta in equation (6), the {\it basic} 
equal-time (anti)commutators for the Lagrangian  density  ${\cal L}_{(B)}$ [cf. Eq. (1)] are 
\begin{eqnarray}\label{9}
&& \big \{ C^a (\vec{x}, \, t), \; \; \dot {\bar C} (\vec{y},  \, t) \big \} =  \delta^{ab}\, \delta^{(D-1)} (\vec{x} - \vec{y}), \nonumber\\
&& \big \{ \bar C^a (\vec{x}, \, t), \; \; (D_0 C)^b (\vec{y},   \, t) \big \} = - \, \delta^{ab}\, \delta^{(D-1)} (\vec{x} - \vec{y}), \nonumber\\
&& \big [ A_{0}^a (\vec{x}, \, t), \; \; \,\, B^b (\vec{y},  \, t) \big ]= i \, \delta^{ab}\, \delta^{(D-1)} (\vec{x} - \vec{y}), \nonumber\\
&& \big [ A_{i}^a (\vec{x}, \, t), \; \; \, \,F^{0jb} (\vec{y},  \, t) \big ] = -\,i \, \delta^{ab}\,\delta_i^{j}\, \delta^{(D-1)} (\vec{x} - \vec{y}),
\end{eqnarray}
and the rest of {\it all} the canonical 
(anti)commutators are {\it zero} (for the Lagrangian  density  ${\cal L}_{(B)}$). In other words, we have already derived
the covariant canonical quantization of the BRST-quantized version of our D-dimensional non-Abelian 1-form gauge theory that is described by the 
Lagrangian  density  ${\cal L}_{(B)}$. Now we concentrate on the Lagrangian  density  ${\cal L}_{(\bar B)}$.  Taking the help
from (7), we obtain the analogues of the canonical 
equal-time (anti)commutators of equation (9), for the Lagrangian density ${\cal L}_{(\bar B)}$ [cf. Eq. (1)], as:
\begin{eqnarray}\label{10}
&& \big \{ C^a (\vec{x}, \, t), \; \; (D_0 {\bar C})^b (\vec{y},  \, t) \big \} =  \delta^{ab}\, \delta^{(D-1)} (\vec{x} - \vec{y}), \nonumber\\
&& \big \{ \bar C^a (\vec{x}, \, t), \; \; \,\dot C^b (\vec{y},   \, t) \big \} = - \, \delta^{ab}\, \delta^{(D-1)} (\vec{x} - \vec{y}), \nonumber\\
&& \big [ A_{0}^a (\vec{x}, \, t), \; \; \, \,\bar B^b (\vec{y},  \, t) \big ]= -\, i \, \delta^{ab}\, \delta^{(D-1)} (\vec{x} - \vec{y}), \nonumber\\
&& \big [ A_{i}^a (\vec{x}, \, t), \; \; \, \, F^{0jb} (\vec{y},   \, t) \big ] = -\,i \, \delta^{ab}\,\delta_i^{j}\, \delta^{(D-1)} (\vec{x} - \vec{y}).
\end{eqnarray}
The rest of {\it all} the possible (anti)commutators are {\it zero}. In other words, we have already obtained 
the BRST-quantization (i.e. covariant canonical quantization) of 
our D-dimensional non-Abelian 1-form gauge theory that is described by the Lagrangian density ${\cal L}_{(\bar B)}$.

We wrap-up our present section with the final (and quite important) remarks that 
the {\it basic} canonical (anti)commutators for (i)  the {\it perfectly} anti-BRST invariant Lagrangian density ${\cal L}_{(\bar B)}$,
and (ii) the {\it perfectly} BRST invariant Lagrangian density
 ${\cal L}_{(B)}$, can be utilized in the proof that
the Noether (anti-)BRST charges (5) are the {\it true} generators\footnote{It is pretty straightforward to corroborate this statement if we use
the expressions for the Noether conserved (anti-)BRST charges [cf. Eqs. (47),(32) below] that have been expressed in terms of the canonical 
conjugate momenta [cf. Eqs. (6),(7)]
and exploit the general form of the {\it basic} canonical (ant)commutators from equation (8). For readers' convenience, we have worked out a few 
examples in our Appendix D.}
 for the nilpotent (anti-)BRST symmetry transformations [cf. Eq. (2)]
for the {\it basic} fields $A_\mu, \; C, \; \bar C$.  
To corroborate this statement,
we note that the following {\it general} mathematical relationship
\begin{eqnarray}\label{11}
s_r \, \Phi (\vec{x}, \, t) = -\, i\,
 \big [ \Phi (\vec{x}, \, t), \; \; Q_r \big ]_{(\pm)}, \qquad r = b, \, ab, \qquad  \Phi = A_\mu, \; C, \; \bar C,
\end{eqnarray}
leads to the derivations of the nilpotent 
(anti-)BRST symmetry transformations where the conserved Noether (anti-)BRST charges [cf. Eq. (5)] are the generators. In the above equation (11),
the subscripts $(\pm)$ on the square bracket [on the r.h.s. of equation (11)] correspond to the (anti)commutator for a given 
generic field $\Phi$ of our BRST-quantized theory being fermionic/bosonic in nature. It is worthwhile to mention, at this juncture, that the 
{\it other} (anti-)BRST symmetry transformations in (2), besides the basic fields $A_\mu, \; C, \; \bar C $ of our theory, have been obtained
from the requirements of (i) the absolute anticommutativity property (i.e. $\{ s_{ab}, \; s_{b} \} = 0 $) between the 
(anti-)BRST symmetry transformation operators $s_{(a)b} $, (ii) the off-shell nilpotency  property (i.e. $s_{(a)b}^2 = 0 $) of the transformation
operators $s_{(a)b} $,
and (iii) the (anti-)BRST invariance of the CF-condition [i.e.  $s_{(a)b} [B + \bar B + (\bar C \times C)] = 0 $]. As a side remark,
we would like to point out that the off-shell nilpotent (anti-)BRST symmetry transformations $s_{ab}\; F_{\mu\nu}      = i \,(F_{\mu\nu}\times\bar C) $ and 
$s_{b}\; F_{\mu\nu}      = i \,(F_{\mu\nu}\times C) $ in equation (2) have been derived from the {\it basic}  
off-shell nilpotent (anti-)BRST symmetry transformations (i.e. $s_{ab}\; A_\mu = D_\mu\bar C $ and $s_{b}\; A_\mu= D_\mu  C $)
on the gauge field, respectively, where we have taken into considerations
(i) the definition of the antisymmetric field-strength tensor: $F_{\mu\nu} = \partial_\mu\, A_\nu - \partial_\nu\, A_\mu + i\, (A_\mu \times A_\nu) $
for the non-Abelian 1-form gauge field $A_\mu = A_\mu^a \, T^a$, and (ii) the observations: $[D_\mu, \; D_\nu ]\, \bar C = i\, (F_{\mu\nu} \times \bar C)$
and $[D_\mu, \; D_\nu ]\,  C = i\, (F_{\mu\nu} \times  C)$, respectively, where the covariant derivatives on the ghost and anti-ghost fields:
$D_\mu C = \partial_\mu C + i \, (A_\mu \times C) $  and  $D_\mu \bar C = \partial_\mu \bar C + i \, (A_\mu \times \bar C) $ are in the adjoint 
representations of our chosen SU(N) Lie group and associated SU(N) Lie algebra, respectively.\\


\section{Noether Charges and Their Nilpotency Property}

Our present section is divided into {\it two} subsections. In Subsec. 4.1, we discuss the nilpotency  property of the 
{\it Noether} conserved (anti-)BRST charges
taking into account (i) the continuous {\it off-shell} nilpotent (anti-)BRST symmetry transformations (2), and (ii) their generators as the conserved
(anti-)BRST charges $Q_{(a)b} $ [cf. Eq. (5)] where we demonstrate explicitly that {\it these} charges are neither (anti-)BRST invariant nor they are
nilpotent of order two.
Our Subsec. 4.2 covers the theoretical contents that are connected with the use of (i) the partial integration along with 
the Gauss divergence theorem, and (ii) the appropriate EL-EoM at suitable places, to prove
the nilpotency property of the Noether conserved (anti-)BRST charges $Q_{(a)b} $. \\


\subsection{ Nilpotency Property: Off-Shell Nilpotent Symmetries}

The central theme of our present subsection is to demonstrate that the Noether conserved (anti-)BRST charges $Q_{(a)b} $
are {\it not} nilpotent of order two (i.e. $Q_{(a)b}^2 \neq 0 $) when we utilize the relationship (11) by taking into account the off-shell 
nilpotent {\it continuous} symmetry transformations (2) and the  Noether conserved (anti-)BRST charges (5) as {\it their} generators.
To  corroborate this statement, we use
the general relationship of equation (11), with the identification(s): $\Phi = Q_{(a)b} $, which leads to the following  explicit expressions
\begin{eqnarray}\label{12}
&&s_{ab}\, Q_{ab} = -\, i\,
 \big \{ Q_{ab}, \;  Q_{ab} \big \} \equiv -\, i\, \int d^{D-1} x\, \Big[\, \frac{1}{2}\, \dot {\bar B} \cdot (\bar C \times \bar C)
+ (F^{0i} \times \bar C) \cdot D_i\, \bar C   \Big] \neq 0, \nonumber\\
&&s_{b}\, Q_{b} = = -\, i\,
 \big \{ Q_{b}, \; \; Q_{b} \big \} \equiv i\, \int d^{D-1} x\, \Big[ \frac{1}{2}\, \dot{ B} \cdot( C \times C)
- (F^{0i} \times  C) \cdot D_i\, C   \Big] \neq 0,
\end{eqnarray} 
where the extreme r.h.s. of the above integrals are nothing but the 
explicit computation of the l.h.s. (i.e. $s_{ab}\, Q_{ab}$ and $s_{b}\, Q_{b} $) where we have applied the infinitesimal, continuous and off-shell
nilpotent (anti-)BRST transformations (2) {\it directly} on the precise expressions for the Noether 
(anti-)BRST charges (5). The above observations demonstrate that (i) the Noether conserved (anti-)BRST charges (5) are {\it not} invariant 
(i.e. $s_{ab}\, Q_{ab} \neq 0, \; s_b \, Q_b \neq 0$) under
the nilpotent (anti-)BRST symmetry transformations (2), and (ii) these conserved charges are also {\it not} off-shell nilpotent
(i.e. $Q_{(a)b}^2 \neq 0 $). The {\it latter}
conclusion is drawn from the close observations of the anticommutators in equation (12) which imply that: $-\, i\,
 \big \{ Q_{ab}, \;  Q_{ab} \big \} = -\, 2\,i \,Q_{ab}^2 \neq 0 $ and $-\, i\,
 \big \{ Q_{b}, \;  Q_{b} \big \} = -\, 2\,i\, Q_{b}^2 \neq 0 $. As a consequence, the non-invariant (i.e. $s_{ab}\, Q_{ab} \neq 0, \; s_b \, Q_b \neq 0$) 
Noether conserved (anti-)BRST charges $Q_{(a)b} $ [cf. Eq. (5)]  are {\it not}
suitable for the discussion on the physicality criteria (see, e.g. [5,7]). As a side remark, we would like to state that
a careful look at equation (12) demonstrates that, in the 
{\it Abelian} limit [where the gauge group is U(1) without any non-trivial Lie algebraic space], the Noether non-nilpotent [cf. Eq. (12)]
conserved (anti-)BRST charges reduce to the {\it nilpotent} conserved (anti-)BRST charges which are (anti-)BRST {\it invariant}, too.

We wrap-up our present short (but quite important) subsection with a crucial remark that the conserved Noether (anti-)BRST charges
are {\it not} (anti-)BRST invariant (i.e. $s_{ab}\, Q_{ab} \neq 0, \; s_b \, Q_b \neq 0$). Hence, 
it is clear that (i) they are {\it not} physical quantities within the framework of BRST formalism, and (ii) they will {\it not} be
useful for the discussions on the physicality criteria (i.e. $Q_{(a)b}\, |phys> = 0 $) where it is demanded that the physical states 
(i.e. $|phys> $), in the {\it total} quantum Hilbert space of states,
 are {\it those} that are annihilated by (i) the conserved, and (ii) the (anti-)BRST invariant, versions of the
 (anti-)BRST charges. Furthermore, we note that the Noether (anti-)BRST charges
are {\it not} nilpotent of order two [cf. Eq. (12)] w.r.t. the off-shell nilpotent (i.e. $s_{(a)b}^2 = 0 $)
(anti-)BRST symmetry transformation operators $s_{(a)b} $.  However, in the Abelian limit 
where the SU(N) gauge group reduces to the  U(1) gauge group, {\it these} (anti-)BRST 
charges take the forms that are conserved as well as (anti-)BRST invariant. As far as our observations in equation (12) are concerned,
we are allowed to use (i) the EL-EoM at appropriate places inside the integrals, and (ii) the Gauss divergence theorem (due to which the volume integral 
gets converted into the surface integral and all the physical fields vanish off as $x \to \pm\, \infty $).\\


\subsection{ Nilpotency Property: On-Shell Conditions}

The key purpose of our present subsection is to demonstrate that if we use (i)  the appropriate EL-EoM at suitable places, 
and (ii) the Gauss divergence theorem,
the conserved Noether (anti-)BRST charges [cf. Eqs. (5),(12)] become  the (anti-)BRST invariant quantity and 
 nilpotent of order two. To corroborate this statement, first of all, let us focus on the expression for $s_b Q_b$ in equation (12)
and re-express it in its {\it explicit} form as: 
\begin{eqnarray}\label{13}
s_{b}\, Q_{b} = \, i\,
\int d^{D-1} x\, \Big[ \dfrac{1}{2}\, \dot{ B} \cdot( C \times C)
- (F^{0i} \times  C) \cdot \partial_i\, C  - i \, (F^{0i} \times  C) \cdot (A_i \times C) \Big].
\end{eqnarray} 
The {\it second} term in the above expression can be re-written in the following form
\begin{eqnarray}\label{14}
 - \, i\, (F^{0i} \times  C) \cdot \partial_i\, C  =  \partial_i \big[ -\,i\,F^{0i} \cdot (C \times C) \big ] + i\, \partial_i F^{0i} \cdot (C \times C)
+ \dfrac{i}{2} \,F^{0i} \cdot \partial_i\, (C \times C), 
\end{eqnarray} 
where we have used the usual convention of the dot and cross products in the Lie algebraic space.
The above quantity can be simplified and re-expressed as 
\begin{eqnarray}\label{15}
 - \dfrac{i}{2}\, \partial_i \big[ F^{0i} \cdot (C \times C) \big ] 
+ \dfrac{i}{2} \,(\partial_i F^{0i}) \cdot (C \times C), 
\end{eqnarray}
where we have used the standard mathematical tricks to re-express the {\it last} term 
of equation (14) as: $\partial_i [\frac{i}{2}\, F^{0i} \cdot (C \times C) ] - \frac{i}{2} \, (\partial_i F^{0i}) \cdot (C \times C) $.
Thus, we can write the expression for $s_b \,Q_b $ of
equation (13), in its modified form,  as follows:
\begin{eqnarray}\label{16}
s_{b}\, Q_{b} &=& \, i\,
\int d^{D-1} x\, \Big[ \frac{1}{2}\, \dot{ B} \cdot( C \times C)
- \dfrac{1}{2}\, \partial_i \big[ F^{0i} \cdot (C \times C) \big ] 
+ \dfrac{1}{2} \,(\partial_i F^{0i}) \cdot (C \times C) \nonumber\\
&-& i \, (F^{0i} \times  C) \cdot (A_i \times C) \Big].
\end{eqnarray}
At this juncture, we concentrate on the {\it last} term of the above integral that is located inside the square bracket. In other words, 
we do {\it not} take care of the overall $i$ factor that is present outside the integral. To be precise, the last term
[i.e. $- i \, (F^{0i} \times  C) \cdot (A_i \times C) $]  can be re-arranged, using the {\it usual} rules of the dot and cross products, as follows:  
\begin{eqnarray}\label{17}
- i \, (F^{0i} \times  C) \cdot (A_i \times C)  = + i \, (F^{0i} \times  C) \cdot (A_i \times C) + i\, (A_i \times F^{0i}) \cdot (C \times C), 
\end{eqnarray}
where we have used (i) the anticommutativity property (i.e. $C^a\, C^b + C^b \, C^a = 0 $) of the ghost fields $C^a$, and (ii) the 
explicit relationship amongst the structure constants that emerges out due to the validity of the Jacobi identity in the Lie algebraic space. 
The {\it latter} can be expressed explicitly, in the mathematical language, as follows: 
\begin{eqnarray}\label{18}
f^{abc}\; f^{cmn} \, + \, f^{bmc}\; f^{can} \, + \, f^{mac}\; f^{cbn}  = 0. 
\end{eqnarray}
A close look at equation (17) implies that we have the following equality, namely;
\begin{eqnarray}\label{19}
- i \, (F^{0i} \times  C) \cdot (A_i \times C)  = + \dfrac{i}{2}\; (A_i \times F^{0i}) \cdot (C \times C). 
\end{eqnarray}
The substitution of (19) into the expression $s_{b}\, Q_{b} $ [cf. Eq. (16)] leads to the following
\begin{eqnarray}\label{20}
s_{b}\, Q_{b} = \, i\,
\int d^{D-1} x\, \Big[ \frac{1}{2}\, \dot{ B} \cdot( C \times C)
- \dfrac{1}{2}\, \partial_i \big[ F^{0i} \cdot (C \times C) \big ] 
+ \dfrac{1}{2} \,(D_i F^{0i}) \cdot (C \times C) \Big].
\end{eqnarray}
where the expression for the covariant derivative on $F^{0i} $ (in the adjoint representation of the SU(N) Lie algebraic space)
 is: $D_i F^{0i} = \partial_i \, F^{0i} + i\, (A_i \times F^{0i})$. At this point of discussion, 
we invoke the validity of the EL-EoM: $D_\mu\, F^{\mu\nu} - \partial^\nu\, B - (\partial^\nu\, \bar C \times C) = 0$ which is 
derived from the {\it perfectly} BRST-invariant Lagrangian density ${\cal L}_{(B)} $ w.r.t. the gauge field $A_\mu = A_\mu^a \, T^a$. From this
equation, it is straightforward to check that we have the following equality
\begin{eqnarray}\label{21}
(D_i F^{i0}) \cdot (C \times C) = \dot B \cdot (C \times C) \;\; \Longrightarrow \;\; (D_i F^{0i}) \cdot (C \times C) = -\, \dot B \cdot (C \times C),
\end{eqnarray}
where we have (i) made the choice $\nu = 0 $ in the EL-EoM: $D_\mu\, F^{\mu\nu} - \partial^\nu\, B - (\partial^\nu\, \bar C \times C) = 0$, and
(ii) taken into account the input: $C \times (C \times C) \equiv - \, (C \times C) \times C = 0 $. The {\it latter} is required in 
the proof that $(\dot {\bar C} \times C) \cdot (C \times C) = 0 $ 
where (it goes without saying that)  the beauty of the relationship (18) is also exploited\footnote{It is interesting to point out that, using 
the relationship in (18),
the explicit computation establishes that: $(\dot {\bar C} \times C) \cdot (C \times C) = \dot {\bar C} \cdot [C \times (C \times C)] 
+ \dot {\bar C} \cdot [(C \times C) \times C]$. The r.h.s. of this relationship becomes {\it zero} due to: $ [C \times (C \times C)] = -\,
[(C \times C) \times C]$ which emeges out from the rules of the cross product between a fermionic vector $C^a$ and a bosonic vector $C \times C)^a $
in the SU(N) Lie algebraic space.}. Thus, ultimately, we have been able to prove
that, if we use the EL-EoM, we obtain the {\it final} expression of $s_{b}\, Q_{b} $ as follows:
\begin{eqnarray}\label{22}
s_{b}\, Q_{b} = -\, \dfrac{i}{2}\,
\int d^{D-1} x\,\; \partial_i \big[ F^{0i} \cdot (C \times C) \big ] \equiv -\, i\, \big \{Q_b, \; Q_b \big \}. 
\end{eqnarray}
It is evident that the above integral goes to zero due to the application of the Gauss divergence theorem (because all the 
relevant physical fields 
vanish off as $x \to \pm \, \infty $). In other words, we have shown that the Noether BRST charge $Q_b$ [cf. Eq. (5)] is found (i) to be a
BRST-invariant (i.e. $s_{b}\, Q_{b} = 0 $) quantity, and
(ii) nilpotent (i.e. $Q_b^2 = 0$)  
of order two, provided we utilize the theoretical strength of (i) the Gauss divergence theorem, and (ii) the 
appropriate {\it on-shell} condition (i.e. the EL-EoM w.r.t. the gauge field $A_\mu$).

We end this subsection with a brief discussion on the proof that the Noether anti-BRST charge $Q_{ab}$ [cf. Eq. (5)] is found to be anti-BRST invariant
(i.e. $s_{ab}\, Q_{ab} = 0 $) as well as nilpotent (i.e. $Q_{ab}^2 = \frac{1}{2}\, \{Q_{ab}, \; Q_{ab} \} = 0 $) of order two
provided we (i) use the appropriate EL-EoM at suitable places, and (ii) exploit the theoretical strength of the Gauss divergence theorem. Toward this goal
in mind, first of all, we note that the explicit expression for $s_{ab}\,Q_{ab} $ [cf. Eq. (12)], in its full blaze of glory,  is as follows:
\begin{eqnarray}\label{23}
s_{ab}\, Q_{ab} =  -\, i\, \int d^{D-1} x\, \Big[\, \frac{1}{2}\, \dot {\bar B} \cdot (\bar C \times \bar C)
+ (F^{0i} \times \bar C) \cdot [\partial _i\, \bar C + i\, (A_i \times \bar C)]  \Big].
\end{eqnarray} 
The {\it second} term of the above expression can be re-written as 
\begin{eqnarray}\label{24}
\big (F^{0i} \times \bar C \big) \cdot \partial _i\, \bar C = \dfrac{1}{2} \, \partial_i \big [ F^{0i} \cdot (\bar C \times \bar C)\big]
- \dfrac{1}{2} \, (\partial_i \, F^{0i})  \cdot (\bar C \times \bar C),
\end{eqnarray}
where we have (i) ignored the overall $(-\,i)$ factor that is located outside the integral, (ii) used the usual rules 
for the dot and cross products in the Lie algebraic space, and (iii) fellowed the similar kinds of theoretical tricks as in the 
context of the computation of the $s_b Q_b $ for the Noether conserved BRST charge [cf. Eqs. (14),(15)]. The {\it third} term of equation (23)
can be expressed as follows 
\begin{eqnarray}\label{25}
i\, (F^{0i} \times \bar  C) \cdot (A_i \times \bar C)  = -\,\dfrac{i}{2}\, (A_i \times F^{0i}) \cdot (\bar C \times \bar C), 
\end{eqnarray}
where we have ignored the overall ($-\, i $) factor that is present outside the integral. Taking into account the expressions in (24) and (25),
we can express $s_{ab}\, Q_{ab} $ [cf. Eq. (23)] as:
\begin{eqnarray}\label{26}
s_{ab}\, Q_{ab} = - \, i\,
\int d^{D-1} x\, \Big[ \dfrac{1}{2}\, \dot{ \bar B} \cdot (\bar C \times \bar C)
+ \dfrac{1}{2}\, \partial_i \big[ F^{0i} \cdot (\bar C \times \bar C) \big ] 
- \dfrac{1}{2} \,(D_i F^{0i}) \cdot (\bar C \times \bar C) \Big].
\end{eqnarray}
At this juncture, we invoke the EL-EoM: $D_\mu\, F^{\mu\nu} + \partial^\nu\, \bar B + (\bar C \times \partial^\nu C) = 0$ which is derived from the
{\it perfectly} anti-BRST invariant Lagrangian density ${\cal L}_{(\bar B)} $ w.r.t. the gauge field $A_\mu = A_\mu^a \, T^a$. Taking into account
the choice $\nu = 0 $ in the above EL-EoM, we obtain the following very {\it useful} relationship, namely;
\begin{eqnarray}\label{27}
(D_i F^{0i}) \cdot (\bar C \times \bar C) = \dot {\bar B} \cdot (\bar C \times \bar C),
\end{eqnarray}
where we have used (i) the usual rules of the dot and cross products in the Lie algebraic space, and (ii) the input:
 $\bar C \times (\bar C \times \bar C) \equiv - \, (\bar C \times \bar C) \times \bar C = 0 $. Thus, ultimately, we obtain the
 {\it final} expression for $s_{ab}\, Q_{ab} $ [cf. Eq. (26)] as follows
\begin{eqnarray}\label{28}
s_{ab}\, Q_{ab} = -\, \dfrac{i}{2}\,
\int d^{D-1} x\,\; \partial_i \big[ F^{0i} \cdot (\bar C \times \bar C) \big ] \equiv -\, i\, \big \{Q_{ab}, \; Q_{ab} \big \}, 
\end{eqnarray}
which vanishes off due to the application of the Gauss divergence theorem thereby proving that the Noether anti-BRST charge
[cf. Eq. (5)] is an anti-BRST invariant (i.e. $s_{ab}\, Q_{ab} = 0 $) quantity and it is is also nilpotent 
(i.e. $Q_{ab}^2 = 0 $) of order two. In other words, the key properties of the (i) anti-BRST invariance, and (ii) nilpotency property
owe their origin to the above conditions (i.e. the Gauss divergence theorem along with the EL-EoM). We would like to lay emphasis on the fact
that, while applying the Gauss divergence theorem,  we have assumed that there is {\it no} non-trivial topology at the boundary.\\


\section{(Anti-)BRST Invariant Versions of the Conserved (Anti-)BRST Charges and the Physicality Criteria}

Our present section is divided into {\it three} subsections. In Subsec. 5.1, we precisely derive the 
(anti-)BRST invariant versions of the conserved (anti-)BRST charges
in a consistent manner where (i) the off-shell nilpotent (anti-)BRST symmetry transformations, and (ii) the appropriate EL-EoM at suitable places,
are given utmost importance (see, e.g. [6]). The subject matter of our 
Subsec. 5.2 is connected with the proof of the (anti-)BRST invariance (i.e. $s_{ab} Q_{AB} = 0, \; 
s_{b} Q_{B} = 0 $) of the consistently {\it modified} versions $Q_{(A)B} $  by taking into account the {\it basic} canonical (anti)commutators [cf. Eq. (8)].
Our subsection 5.3 is devoted to the discussion on the physicality criteria w.r.t. (i) the Noether conserved 
charges $Q_{(a)b} $ which are {\it not} invariant under the off-shell nilpotent (anti-)BRST symmetry transformations (2), and (ii) the consistently
modified versions of the (anti-)BRST charges $Q_{(A)B} $ which are {\it invariant} under the off-shell nilpotent (anti-)BRST transformations (2).


\subsection{ (Anti-)BRST Invariant Charges $Q_{(A)B}$: Proof Using Gauss's Divergence Theorem, EL-EoMs and Symmetries}

We have already checked that the Noether conserved (anti-)BRST charges [cf. Eq. (5)] are {\it not} (anti-)BRST invariant 
quantities [cf. Eq. (12)] under the off-shell nilpotent (i.e. $s_{(a)b}^2 = 0 $)
(anti-)BRST symmetry transformations $s_{(a)b} $ that have been listed in equation (2). 
However, we can obtain, in a consistent manner,  the (anti-)BRST invariant (i.e. $s_{ab} Q_{AB} = 0, \; s_b Q_B = 0 $) versions of the 
conserved (anti-)BRST charges $Q_{(A)B} $ from the 
Noether conserved (anti-)BRST charges $Q_{(a)b}$ by exploiting the theoretical strength of 
(i) the partial integration along with the Gauss divergence theorem where the total space derivative term in the volume integral 
 gets converted into the surface integral where all the physical fields 
(that are present  in the argument of the total space derivative term) tend to go to infinity ($x \to \pm \, \infty $) 
and, as a consequence, vanish off (see, e.g. [6]), and
(ii) the EL-EoM w.r.t to the gauge field $A_\mu = A_\mu^a T^a$ from the {\it perfectly} BRST and anti-BRST  invariant Lagrangian  densities
 ${\cal L}_{(B)}$ and  ${\cal L}_{(B)}$, respectively. To be precise, the explicit forms of these EL-EoM 
w.r.t. the gauge field of our BRST-quantized theory are as follows:
\begin{eqnarray}\label{29}
D_\mu\, F^{\mu\nu} - \partial^\nu\, B - (\partial^\nu\, \bar C \times C) = 0, \qquad 
D_\mu\, F^{\mu\nu} + \partial^\nu\, \bar B + (\bar C \times \partial^\nu\, C) = 0.
\end{eqnarray}
The {\it final} forms of the  (anti-)BRST invariant (i.e. $s_{ab} Q_{AB} = 0,  s_b Q_B = 0 $) {\it modified} versions 
$Q_{(A)B}$ (that are derived from 
conserved {\it Noether} (anti-)BRST charges $Q_{(a)b} $) are 
\begin{eqnarray}\label{30}
Q_{ab} \rightarrow Q_{AB} &=& \int d^{D - 1}\,x\,\Big[\dot{\bar B}\cdot {\bar C} - {\bar B}\cdot D_0\,{\bar C}
+ \frac{1}{2}\,{\dot C} \cdot ({\bar C} \times {\bar C}) \Big], \nonumber\\
Q_{b} \rightarrow Q_{B} &=& \int d^{D - 1}\,x\,\Big[{B}\cdot D_0\,{C} - \dot {B}\cdot {C}
- \frac{1}{2}\,\dot{\bar C} \cdot({C} \times {C}) \Big],
\end{eqnarray}
which have been precisely obtained in our earlier work (see, e.g. [6] for details).
We have claimed, in our earlier work [6], that {\it these} (anti-)BRST invariant charges $Q_{(A)B} $ are 
{\it also} off-shell nilpotent (i.e. $Q_{(A)B}^2 = 0 $)
by using the formula: $s_{ab} Q_{AB} = -\, i\,  \big \{ Q_{AB}, \;  Q_{AB} \big \} = 0 $ and $s_{b} Q_{B} = -\, i\,  \big \{ Q_{B}, \;  Q_{B} \big \} = 0 $
which is {\it not} correct in view of our equation (11) where we have claimed that the conserved Noether (anti-)BRST charges 
$Q_{(a)b} $ are the {\it true}
generators for the off-shell nilpotent (i.e. $s_{(a)b}^2 = 0 $) (anti-)BRST  transformations $s_{(a)b}$.

We wrap-up our present short (but quite useful) subsection with the remarks that our
claims, in the earlier work [6], are correct as far as the (anti-)BRST invariance (i.e. $s_{ab} Q_{AB} = 0, \; s_b Q_B = 0 $) of the conserved (anti-)BRST
charges $Q_{(A)B} $ are concerned. Since we have derived the {\it modified} versions of the (anti-)BRST charges $Q_{(A)B}$ from the Noether {\it conserved}
(anti-)BRST charges $Q_{(a)b}$ by exploiting (i) the partial integration along with the Gauss divergence theorem, and (ii) the appropriate 
EL-EoM at suitable places, the {\it modified} versions of the (anti-)BRST charges $Q_{(A)B}$ are also {\it conserved} as per the basic tenets of 
the Noether theorem. We would like 
to add that {\it both} types of conserved (anti-)BRST charges (i.e. $Q_{(a)b} $ and $Q_{(A)B} $) are useful in their {\it own} right (within the
framework of BRST formalism). Whereas the Noether conserved (anti-)BRST charges $Q_{(a)b} $ are the generators [cf. Eq. (11)] for the 
off-shell nilpotent versions of the 
(anti-)BRST symmetry transformations (2), the conserved and  (anti-)BRST invariant charges  $Q_{(A)B} $ are {\it physical} quantities\footnote{ Within
the framework of BRST formalism, the quantities that are BRST as well as anti-BRST invariant, are really {\it physical} quantities. Hence, the
conserved and (anti-)BRST invariant versions of the (anti-)BRST charges $Q_{(A)B} $ are used in the physicality criteria (cf. Subsec. 5.3 below).}  
w.r.t. to the off-shell nilpotent (anti-)BRST symmetry transformations (2) and, hence, they are quite handy in the discussion on 
the physicality criteria (i.e. $Q_{(A)B}\, |phys> = 0 $) where the physical states (i.e. $|phys> $), existing in the {\it total} 
quantum Hilbert space of states, are {\it those} that are annihilated by the (anti-)BRST {\it invariant}
and  {\it conserved} versions of the (anti-)BRST charges  $Q_{(A)B}$. In fact, we have
compared and contrasted the roles of the Noether conserved charges $Q_{(a)b} $ and {\it modified} versions of the (anti-)BRST
charges $Q_{(A)B}$ in the context of the physicality criteria (cf. Subsec. 5.3 below for details).\\



\subsection{(Anti-)BRST Invariant Conserved Charges $Q_{(A)B}$: Proof Using the Basic Canonical Brackets}

The central objective of our present subsection is to show that the (anti-)BRST charges  $Q_{(A)B} $ are the {\it true} physical quantities
(within the framework of BRST formalism) because they remain invariant under the off-shell nilpotent (anti-)BRST symmetry
transformations (2) when we use (i) the relationship (11), and (ii) the basic canonical (anti)commutators (8) [and/or
their off-shoots that have been listed in equations (9) and (10)]. To corroborate this assertion, we observe (from the relationship:
$s_b Q_B = -\,i\, \{Q_B, \; Q_b \}$)  the following
\begin{eqnarray}\label{31}
\Big  \{ Q_B, \; Q_b \Big \} &=& \int \int \, d^{D-1} x\; d^{D-1} y \; 
\Big \{ \Big (i \, \Pi^{0}_{(A)} \cdot \Pi_{(\bar C)} - \dot B \cdot  C + \dfrac{i}{2} \, \Pi_{(C)} \cdot (C \times C) \Big ) (\vec{x}, t), \;
\nonumber\\
&& \Big (i \, \Pi^{0}_{(A)} \cdot \Pi_{(\bar C)} + \Pi^{i}_{(A)} \cdot D_i C - \dfrac{i}{2} \, \Pi_{(C)} \cdot (C \times C) \Big ) (\vec{y}, t)\; \Big \},
\end{eqnarray}
where we have used the precise expression for $Q_B$ and $Q_b$ 
in terms of the canonical conjugate momenta [cf. Eq. (6)] that have been derived from the {\it perfectly} BRST invariant Lagrangian density 
${\cal L}_{(B)}$ [cf. Eq. (1)]. In other words, we have the following explicit expressions:
\begin{eqnarray}\label{32}
Q_B  &=&  \int \, d^{D-1} x\;  
\Big (i \, \Pi^{0}_{(A)} \cdot \Pi_{(\bar C)} - \dot B \cdot  C + \dfrac{i}{2} \, \Pi_{(C)} \cdot (C \times C) \Big ) (\vec{x}, t), \;
\nonumber\\
Q_b &=& \int \, d^{D-1} x\;
 \Big (i \, \Pi^{0}_{(A)} \cdot \Pi_{(\bar C)} + \Pi^{i}_{(A)} \cdot D_i C - \dfrac{i}{2} \, \Pi_{(C)} \cdot (C \times C) \Big ) (\vec{x}, t).
\end{eqnarray}
It is crystal clear that we have taken into account the expressions for (i) the BRST invariant (i.e. $s_b Q_B = 0$) conserved charge 
$Q_B$ from equation (30), and (ii) the Noether conserved BRST charge $Q_b$ from equation (5). On top of it, we have
re-expressed {\it them} in terms of the canonical conjugate momenta [cf. Eq. (6)].
The existing non-zero anticommutators, from the above full form of the expression for $\{Q_B, \; Q_b \} $ [cf. Eq. (31)], are as follows
\begin{eqnarray}\label{33}
&& \Big  \{ Q_B, \; Q_b \Big \} =
\dfrac{i}{2}\, \int \int \, d^{D-1} x\; d^{D-1} y \;
\Big \{ \big (\dot B \cdot C \big ) (\vec{x}, t), \; \Big (\Pi_{(C)} \cdot (C \times C) \Big ) (\vec{y}, t) \Big \} \nonumber\\
&+& \dfrac{i}{2}\, \int \int \, d^{D-1} x\; d^{D-1} y \;
\Big \{ \Big (\Pi_{(C)} \cdot (C \times C) \Big ) (\vec{x}, t), \; \Big (\Pi^{i}_{(A)} \cdot D_i C \Big ) (\vec{y}, t) \Big \} \nonumber\\
&+& \dfrac{1}{4}\, \int \int \, d^{D-1} x\; d^{D-1} y \,
\Big  \{ \Big (\Pi_{(C)} \cdot (C \times C) \Big ) (\vec{x}, t), \, \Big (\Pi_{(C)} \cdot (C \times C) \Big ) (\vec{y}, t) \Big \},
\end{eqnarray}
where it is straightforward to note that the term ``$i\,\Pi^{0}_{(A)} \cdot \Pi_{(\bar C)} $'' anticommutes with {\it all}
the terms (of $Q_b$) because their canonically conjugate fields $A_0$ and $\bar C$ are {\it not} present in the
full expression for $Q_b$ [cf. Eq. (32)]. It is worthwhile to mention that the {\it last} anticommutator in the above equation (33)
is {\it zero} because of the observation that: $C \times (C \times C) = -\, (C \times C) \times C = 0$ which appears in
the {\it final} expression for the {\it last} integral in (33) with the integrand as: $\Pi_{(C)} \cdot [C \times (C \times C) ] $.
We very briefly sketch here a few key steps to corroborate our assertion.
First of all, we focus on the anticommutator that is present in the last integral of equation (33)
where the {\it existing} terms (from the following anticommutator), namely;
\begin{eqnarray}\label{34}
 \dfrac{1}{4}\, \Big \{ \Big (\Pi_{(C)} \cdot (C \times C) \Big ) (\vec{x}, t), \;\; \Big(\Pi_{(C)} \cdot (C \times C) \Big ) (\vec{y}, t) \Big \} ,
\end{eqnarray}
are the following
\begin{eqnarray}\label{35}
&+& \dfrac{1}{4}\, \Pi_{(C)}^a (\vec{x}, t) \Big [ \big (C \times C \big )^a (\vec{x}, t), \;\; \big (\Pi_{(C)} \cdot (C \times C) \big ) (\vec{y}, t) \Big ]
\nonumber\\
&+& \dfrac{1}{4}\, \Big \{ \Pi_{(C)}^a (\vec{x}, t),\;\; \big (\Pi_{(C)} \cdot (C \times C) \big ) (\vec{y}, t) \Big \} \, (C \times C)^a (\vec{x}, t),
\end{eqnarray}
where we have exploited the suitable rules of the (anti)commutator from our Appendix C.
The {\it first} bracket in the above leads, ultimately,  
to the following {\it non-zero} brackets in terms of the {\it basic} canonical anticommutators, namely;
\begin{eqnarray}\label{36}
&+& \dfrac{1}{4}\, f^{amn} \,\;\Pi_{(C)}^a (\vec{x}, t) \,\; C^m (\vec{x}, t) \;
\Big \{ C^n (\vec{x}, t), \, \Pi^b_{(C)} (\vec{y}, t) \Big \}\, (C \times C)^b (\vec{y}, t)
\nonumber\\
&-& \dfrac{1}{4}\; f^{amn} \,\;\Pi_{(C)}^a (\vec{x}, t) \; \Big \{ C^m (\vec{x}, t), \;\; \Pi^b_{(C)} (\vec{y}, t) \Big \}\, C^n (\vec{x}, t) \,
(C \times C)^b (\vec{y}, t),
\end{eqnarray}
where we have used the standard (anti)commutator rules from our Appendix C.
Using the appropriate basic anticommutator from (8), we obtain the {\it exact} value of the {\it first} term/bracket of equation (35), along
with the integration of the last term of (33), as 
\begin{eqnarray}\label{37}
&+& \dfrac{i}{2}\, f^{amb} \,\;\int d^{D-1} x\; \Big (\Pi_{(C)}^a \, C^m \,(C \times C)^b \Big ) (\vec{x}, t) \nonumber\\
&=& 
+ \dfrac{i}{2}\, \int d^{D-1} x\, \Big (\Pi_{(C)} \cdot [C \times (C \times C)]\Big ) (\vec{x}, t),
\end{eqnarray}
where we have already performed the volume integration over the $y $ variable taking into
account the contribution from the Dirac $\delta$-function. It is straightforward to check that the {\it second} bracket of (35) 
contributes exactly the {\it same} value. Hence, the total sum is {\it twice} of what we have obtained in (37). However, as pointed out earlier,
the {\it net} contribution of the {\it last} term of (33) is {\it zero} due to: $C \times (C \times C) = -\, (C \times C) \times C = 0$.

Let us focus on the {\it first} integral along with the anticommutator on the r.h.s. of the expression $  \{ Q_B, \; Q_b  \} $. Using some of the appropriate
rules of the composite anticommutators from our Appendix C, we find that the {\it first} integral [on the r.h.s. of equation (33)] 
reduces to the following {\it final} form of the expression:
\begin{eqnarray}\label{38}
 \dfrac{i}{2}\, \int \int \, d^{D-1} x\; d^{D-1} y \;
\dot B^a  (\vec{x}, t)\, \Big \{ C^a  (\vec{x}, t),\; \Pi^b_{(C)} (\vec{y}, t) \Big \}\; (C \times C)^b (\vec{y}, t).
\end{eqnarray}
Using the appropriate (i.e. $\{ C^a (\vec{x}, \, t), \; \Pi^b_{(C)} (\vec{y},  \, t) \} = i \, \delta^{ab}\, \delta^{(D-1)} (\vec{x} - \vec{y}) $)
canonical anticommutator from equation (8), it is clear that the final expression for the above integral turns out to be the following
\begin{eqnarray}\label{39}
-\, \dfrac{1}{2}\, \int \, d^{D-1} x\; \big [
\dot B \cdot (C \times C) \big ] \, (\vec{x}, t),
\end{eqnarray}
where we have taken into account the property of the Dirac $\delta$-function in carrying out the volume integral over the $y$ variable in equation (38).

As far as the evaluation of the {\it second} anticommutator (that is present on the r.h.s. of the repression $ \{ Q_B, \; Q_b \} $ [cf. Eq. (33)]
is concerned, we note that it can be expanded into the sum of a set of {\it two} anticommutators as follows
\begin{eqnarray}\label{40}
&& \dfrac{i}{2}\, \int \int \, d^{D-1} x\; d^{D-1} y \;
\Big \{ \Big (\Pi_{(C)} \cdot (C \times C) \Big ) (\vec{x}, t), \; \Big (\Pi^{i}_{(A)} \cdot \partial_i C \Big ) (\vec{y}, t) \Big \} \nonumber\\
&&+ \dfrac{i}{2}\, \int \int \, d^{D-1} x\; d^{D-1} y \;
\Big \{ \Big (\Pi_{(C)} \cdot (C \times C) \Big ) (\vec{x}, t), \; i\, \Big (\Pi^{i}_{(A)} \cdot (A_i \times C) \Big ) (\vec{y}, t) \Big \},
\end{eqnarray}
where we have used the definition of the covariant derivative: $D_i C = \partial_i C + i\, (A_i \times C) $ in the adjoint representation.
The {\it first} of the above integrals can be explicitly expressed as
\begin{eqnarray}\label{41}
\dfrac{i}{2}\, \int \int \, d^{D-1} x\; d^{D-1} y \; \Pi^{ib}_{(A)} (\vec{y}, t) \Big \{ \Pi^a_{(C)} (\vec{x}, t), \; \partial_i C^b (\vec{y}, t) \big \}
\; (C \times C)^a (\vec{x}, t),
\end{eqnarray}
where we have used the appropriate anticommutator rules for the composite operators from our Appendix C.
Using (i) the appropriate basic canonical anticommutator from equation (8), and (ii) the method of partial integration (along with taking care of the
derivative on the Dirac $\delta$-function), we obtain the following expression from the above integral, namely;
\begin{eqnarray}\label{42}
&&+\, \dfrac{1}{2}\, \int \int \, d^{D-1} x\; d^{D-1} y \;\, \partial_i \Pi^{ib}_{(A)} (\vec{y}, t)\; \delta^{ab}\; 
\delta^{D-1} (\vec{x} - \vec{y}) 
\; (C \times C)^a (\vec{x}, t)\nonumber\\
&& \equiv +\, \dfrac{1}{2}\, \int  \, d^{D-1} x \; \Big [\big (\partial_i F^{i0} \big ) \cdot   
\; (C \times C)  \Big ] (\vec{x}, t),
\end{eqnarray}
where we have performed the volume integration over the $y$ variable
and taken into account: $\Pi^{ia}_{(A)} = -\, F^{0ia} \equiv  +\, F^{i0a}$ [cf. Eq. (6)] as well as the property of the Dirac $\delta$-fnction. 
Exploiting the mathematical tricks connected with (i) the volume integral 
over the $y$ variable, (ii) the use of the Dirac $\delta$-function, and (iii) the rules if the dot and cross products in the SU(N) Lie algebraic space,
we find that the {\it second} integral on the r.h.s. of the above equation (40) can be, finally, expressed as follows:
\begin{eqnarray}\label{43}
+\, \dfrac{1}{2}\,  \int \, d^{D-1} x\; \Big [ i\,
(A_i \times F^{i0}) \cdot (C \times C) \Big ](\vec{x}, t).
\end{eqnarray}
Adding the explicit integrals in equations (42) and (43), we obtain the following
\begin{eqnarray}\label{44}
&&+\, \dfrac{1}{2}\,  \int \, d^{D-1} x\; \Big [ \big [\partial_i F^{i0} + i\,
(A_i \times F^{i0}) \big ] \cdot \,(C \times C) \Big ](\vec{x}, t) \nonumber\\
&& \equiv  \dfrac{1}{2}\,  \int \, d^{D-1} x\; \Big [(D_i F^{i0}) 
\cdot \,(C \times C) \Big ](\vec{x}, t),
\end{eqnarray}
where the covariant derivative: $D_i F^{i0} = \partial_i F^{i0} + i\, (A_i \times F^{i0}) $ is in the adjoint representation.
At this stage, we invoke the validity of the EL-EoM w.r.t. the gauge field $A_\mu$ [cf. Eq. (29)] and its ensuing consequence
that leads to: $(D_i F^{i0}) \cdot (C \times C) = \dot B \cdot (C \times C) $ [cf. Eq. (21)]. Taking into account the {\it final} results of
the {\it first} integral [cf. Eq. (39)] and the {\it second} integral [cf. Eq. (44)], we obtain, ultimately, the {\it grand} final result as 
\begin{eqnarray}\label{45}
\Big  \{ Q_B, \; Q_b \Big \} = \dfrac{1}{2}\, \int \, 
d^{D-1} x\; \Big [ -\, \dot B \cdot (C \times C) + \dot B \cdot (C \times C) \Big ] \,(\vec{x}, t) = 0.
\end{eqnarray}
In other words, we have already proven the BRST invariance (i.e. $s_b Q_B = \{ Q_B, \; Q_b \} = 0$) of the 
consistently {\it modified} version of the conserved BRST charge $Q_B$  in the language of the {\it basic} canonical
(anti)commutators that have been listed in equation (8) [and/or their off-shoots in equations (9) and (10)] for the 
Lagrangian densities ${\cal L}_{(B)} $ and ${\cal L}_{(\bar B)} $, respectively.

We end this subsection with a very brief sketch of our discussion on the proof of the anti-BRST invariance: $s_{ab} Q_{AB} = -i\, \{ Q_{AB}, \; Q_{ab} \} = 0$ 
of the {\it modified} version of the conserved anti-BRST charge $Q_{AB}$
taking into account  (i) an appropriate EL-EoM at suitable place, and
(ii) the theoretical strength of the Gauss divergence theorem. The {\it modified} charge $Q_{AB}$
[cf. Eq. (30)]  is, of course, derived form the Noether conserved anti-BRST charge $Q_{ab}$ [cf. Eq. (5)]. 
To prove the anti-BRST invariance 
of the anti-BRST charge $Q_{AB}$, first of all, 
we note that the explicit form of the anticommutator $\{ Q_{AB}, \; Q_{ab} \} $ is  
\begin{eqnarray}\label{46}
\Big  \{ Q_{AB}, \; Q_{ab} \Big \} &=& \int \int \, d^{D-1} x\; d^{D-1} y \; \Big \{  \Big ( \dot {\bar B} \cdot  \bar C 
+\, i \, \Pi^{0}_{(A)} \cdot \Pi_{(C)} + \dfrac{i}{2} \, \Pi_{(\bar C)} \cdot (\bar  C \times \bar C) \Big ) (\vec{x}, t), \;
\nonumber\\
&& \Big (\Pi^{i}_{(A)} \cdot D_i\, \bar C - i \, \Pi^0_{(A)} \cdot \Pi_{(C)}
-\, \frac{i}{2}\, \Pi_{(\bar C)} \cdot (\bar C \times \bar C)  \Big ) (\vec{y}, t)\; \Big \},
\end{eqnarray}
where we have taken the help of the precise expressions $Q_{AB}$ and $Q_{ab}$,  namely;
\begin{eqnarray}\label{47}
Q_{AB} &=&  \int \, d^{D-1} x\;    \Big ( \dot {\bar B} \cdot  \bar C 
+\, i \, \Pi^{0}_{(A)} \cdot \Pi_{(C)} + \dfrac{i}{2} \, \Pi_{(\bar C)} \cdot (\bar  C \times \bar C) \Big ) (\vec{x}, t), 
\nonumber\\
 Q_{ab} &=& \int \, d^{D-1} x\; 
\Big (\Pi^{i}_{(A)} \cdot D_i\, \bar C - i \, \Pi^0_{(A)} \cdot \Pi_{(C)}
-\, \frac{i}{2}\, \Pi_{(\bar C)} \cdot (\bar C \times \bar C)  \Big ) (\vec{x}, t),
\end{eqnarray}
which are nothing but the precise expressions for (i) the anti-BRST invariant (i.e. $s_{ab} Q_{AB} = 0$) 
version of the anti-BRST charge $Q_{AB}$ from 
equation (30), and (ii) the Noether conserved anti-BRST charge $Q_{ab}$ from equation (5) that have been expressed
in terms of the canonical conjugate momenta [cf. Eq. (7)] which have been obtained from the {\it perfectly} anti-BRST invariant
Lagrangian density ${\cal L}_{(\bar B)} $. The 
existing anticommutators, from the above {\it full} form of the 
mathematical expression for $\{Q_{AB}, \; Q_{ab} \} $, are as follows
\begin{eqnarray}\label{48}
&& \Big  \{ Q_{AB}, \; Q_{ab} \Big \} =
-\, \dfrac{i}{2}\, \int \int \, d^{D-1} x\; d^{D-1} y \;
\Big \{ \big (\dot {\bar B} \cdot {\bar C} \big ) (\vec{x}, t), \; \Big (\Pi_{(\bar C)} \cdot (\bar C \times \bar C) \Big ) (\vec{y}, t) \Big \} \nonumber\\
&+& \dfrac{i}{2}\, \int \int \, d^{D-1} x\; d^{D-1} y \;
\Big \{ \Big (\Pi_{(\bar C)} \cdot (\bar C \times \bar C) \Big ) (\vec{x}, t), \; \Big (\Pi^{i}_{(A)} \cdot D_i \bar C \Big ) (\vec{y}, t) \Big \} \nonumber\\
&+& \dfrac{1}{4}\, \int \int \, d^{D-1} x\; d^{D-1} y \,
\Big  \{ \Big (\Pi_{(\bar C)} \cdot (\bar C \times \bar C) \Big ) (\vec{x}, t), \, \Big (\Pi_{(\bar C)} \cdot (\bar C \times \bar C) \Big ) (\vec{y}, t) \Big \},
\end{eqnarray}
where we have taken into account the fact that the term ``$-\, i\, \Pi^{0}_{(A)} \cdot \Pi_{(C)} $'' anticommutes with {\it all} the terms of 
$Q_{ab}$ [cf. Eq. (47)] because there are no canonically conjugate fields $A_0$ and $C$ in the whole expression for $Q_{ab}$. The third anticommutator on 
the r.h.s. of equation (48) would turn out to be {\it zero} due to our observation that: $(\bar C \times \bar C) \times \bar C
= -\, \bar C \times (\bar C \times \bar C) = 0 $. Using exactly the same type of theoretical tricks that 
we have adopted for the proof of the BRST invariance (i.e. $s_b Q_B = -\, i\, \{ Q_B, \; Q_b \} = 0 $)
of the BRST charge $Q_B$, we note that the {\it first} anticommutator on the r.h.s. of equation (48), leads to its final expression as:
\begin{eqnarray}\label{49}
+\, \dfrac{1}{2}\, \int \, d^{D-1} x\; \big [
\dot {\bar B} \cdot (\bar C \times \bar C) \big ] \, (\vec{x}, t).
\end{eqnarray}
On the other hand, the {\it second} anticommutator on the r.h.s. of equation (48), leads to the following form of the
{\it final} integral, namely;
\begin{eqnarray}\label{50}
-\,\dfrac{1}{2}\, \int \, d^{D-1} x\; \Big [(D_i F^{0i}) 
\cdot \,(\bar C \times \bar C) \Big ](\vec{x}, t) \equiv -\, \dfrac{1}{2}\, \int \, d^{D-1} x\; \big [
\dot {\bar B} \cdot (\bar C \times \bar C) \big ] \, (\vec{x}, t),
\end{eqnarray}
where the covariant derivative: $D_i F^{0i} = \partial_i F^{0i} + i\, (A_i \times F^{0i}) $ is in the adjoint representation in the
SU(N) Lie algebraic space. We have taken into account (i) the expression for the canonical conjugate momenta: $\Pi^{ia}_{(A)} = -\, F^{0ia} $,
(ii) used the EL-EoM: $D_\mu F^{\mu\nu} + \partial^\nu \bar B + ({\bar C} \times \partial^\nu  C) = 0 $ 
which is derived from the {\it perfectly} anti-BRST-invariant 
Lagrangian density ${\cal L}_{(\bar B)}$ w.r.t.  the gauge field $A_\mu$ [cf. Eq. (29)], and (iii) chosen $\nu = 0 $ component of
it to prove that $ \dot {\bar B} \cdot (\bar C \times \bar C) = (D_i F^{0i}) \cdot \,(\bar C \times \bar C) $ [cf. Eq. (27)]. 
Finally, we note that the {\it sum} of equations (49) and (50) leads to the proof of
the anti-BRST invariance (i.e.   $s_{ab} Q_{AB} = -\, i\, \{ Q_{AB}, \; Q_{ab} \} = 0 $) of 
the {\it modified} version of the conserved anti-BRST charge $Q_{AB}$ where we have used 
{\it only} the
theoretical strengths of (i) the {\it basic} canonical (anti)commutators (8) [and/or their off-shoots 
that have been expressed in equations (9) and (10)], (ii) the partial integration along with the appropriate EL-EoM at suitable places, and (iii) 
the Gauss divergence theorem. \\

\vskip 0.6cm


\subsection{Physicality Criteria: $Q_{(a)b}$ Versus $Q_{(A)B}$}

A close and careful look at the Noether conserved (anti-)BRST charges [cf. Eq. (5)] shows that every individual term of the Noether
BRST charge $Q_{b} $ carries the effective ghost number equal to $+\, 1$. Furthermore, all the terms in the expression for $Q_b$ are made-up
of (i) the physical fields (carrying the ghost number equal to zero) and a ghost field with the ghost number  $+\, 1$, and/or (ii) only
the combinations of the (anti-)ghost fields [e.g. (1/2) $ \dot {\bar C} \cdot (C \times C) $] with effective ghost number
equal to  $+\, 1$.  In the discussion on the physicality criteria 
 w.r.t. the (anti-)BRST charges, first of all, we invoke the ideas
from the BRST-quantized version of the D-dimensional {\it Abelian} 1-form gauge theory where, right from the beginning, the ghost fields are decoupled from the
physical (i.e. gauge) fields (see, e.g. [20] for details). Hence, the ghost fields (with specific non-zero ghost numbers)
act on the ghost states (i.e. $|ghost> $) and  the physical fields (with ghost number equal to zero) act on the physical
states (i.e. $|phys> $). In the context of the discussion on the BRST-quantized version of the D-dimensional {\it non-Abelian} 1-form gauge
theory, we invoke the {\it same} ideas to discuss the physicality criteria (i.e. $Q_{(a)b} \,|phys> = 0$ and $Q_{(A)B}\, |phys> = 0$)
w.r.t. (i) the Noether conserved (anti-)BRST charges $Q_{(a)b}$,  and (ii) the {\it modified} versions of the 
conserved (anti-) BRST charges $Q_{(A)B}$ (that have been derived  from $Q_{(a)b}$ in a consistent manner). These {\it latter} (anti-)BRST charges
$Q_{(A)B}$ are conserved and (anti-)BRST invariant. Hence, they are 
truly {\it physical} quantities within the framework of BRST formalism.

Let us, first of all, focus on the physicality criterion w.r.t. the Noether conserved BRST charge $Q_b$ [cf. Eq. (5)] and demand
that the physical states (i.e. $|phys> $), existing in the {\it total} quantum Hilbert space of states, are 
{\it those} that are annihilated by the conserved BRST charge $Q_b$. In this context, it is essential to point out that we shall 
demand that {\it all} the individual terms of $Q_b$ will act, separately and independently,  on the physical states (i.e. $|phys> $) and the ghost states (i.e. $|ghost> $). We note that the {\it first} term 
(i.e. $B \cdot D_0 C = B^a (D_0 C)^a  \equiv i\,\Pi^{0}_{(A)} \cdot \Pi_{(\bar C)} $) 
consists of two parts. This {\it first} part $B^a$ happens to be the {\it physical} field which
carries the ghost number equal to {\it zero} and the {\it second} one $(D_0 C)^a $ carries the 
ghost number equal to $ +\, 1$. Hence, the {\it latter} will act on the ghost states and will produce a non-zero result (with eigen value $+\, 1$).
As a consequence, to satisfy the subsidiary (i.e. physicality) criterion w.r.t. the Noether BRST charge (i.e. $Q_{b} \,|phys> = 0$), we have to impose the 
following restriction on the physical states, namely;
\begin{eqnarray}\label{51}
B^a \, |phys> = 0 \;\; \Longrightarrow  \;\; \Pi^{0a}_{(A)} \, |phys> = 0,
\end{eqnarray}
where, as is obvious, we have used: $\Pi^{0a}_{(A)} = B^a $ [cf. Eq. (6)].
In other words, the operator form of the primary constraint (cf. Appendix A for details)
annihilates the physical state (i.e. $|phys> $). This observation is consistent
with the Dirac quantization conditions on the gauge systems that are endowed with the first-class constraints 
in the terminology of Dirac's prescription for the classification scheme of constraints (see, e.g. [11,12]). Now let us focus on the 
{\it second} term of the expression for $Q_b$ [cf. Eq. (5)] which is equal to: $-\, F^{0i} \cdot D_i C = \Pi^{ia}_{(A)} \, (D_i C)^a $. From this
term, it is clear that $(D_i C)^a $ carries the ghost number equal to $+\, 1$. Hence, it will act on the ghost states and produce a {\it non-zero} result.
On the other hand, the term $-\, F^{0ia } =  \Pi^{ia}_{(A)}$ [cf. Eq. (6)]
carries the ghost number equal to {\it zero}. Thus, it is a {\it physical} field and, therefore, it will act on the physical states as:
\begin{eqnarray}\label{52}
-\, F^{0ia}  \, |phys> = 0   \;\; \Longrightarrow  \;\;  \Pi^{ia}_{(A)} \, |phys> = 0.
\end{eqnarray}
However, this result is {\it absurd}  in the sense that $\Pi^{ia}_{(A)} $ is {\it not} a constraint on our theory and it has nothing to do with
the {\it secondary} constraint on our theory that has been discussed in our Appendix A. This {\it wrong} result is obtained 
 because of the fact that the Noether BRST charge $Q_b$ is
{\it not} a physical quantity (within the framework of BRST formalism) because it is {\it not} a BRST-invariant quantity [cf. Eq. (12)]
under the off-shell nilpotent (i.e. $ s_b^2 = 0$) BRST symmetry transformations $s_b$ that have been listed in our equation (2).

Before we end
our discussions on the physicality criteria w.r.t. the conserved Noether (anti-)BRST charges $Q_{(a)b}$, we would like to make
a couple of important remarks. First, we note that the {\it third} term [i.e. (1/2) $\dot {\bar C} \cdot (C \times C) $],
in the expression for the Noether BRST charge $Q_b$ [cf. Eq. (5)],
carries an effective ghost number $+\, 1$. Hence, it acts (i) only on the ghost states, and
(ii) produces a non-zero result. Thus, it will not participate in the physicality criterion: $Q_{b} \,|phys> = 0$ w.r.t. the conserved  Noether 
BRST charge $Q_b$.
Second, we would like to point out that all our discussions, connected with the physicality criterion (i.e. $Q_{b} \,|phys> = 0$)
w.r.t. the BRST charge $Q_b$, will also be {\it true} for the physicality criterion (i.e. $Q_{ab} \,|phys> = 0$) w.r.t. the conserved 
Noether anti-BRST charge $Q_{ab}$  because we shall end-up with the results: $\Pi^{0a}_{(A)} \, |phys> = 0 $ and $\Pi^{ia}_{(A)} \, |phys> = 0 $
which do {\it not} carry any information about the {\it secondary} constraint that
exists on the {\it classical} version  of our D-dimensional non-Abelian 1-form gauge theory (cf. Appendix A for details).

We are now in the position to discuss the physicality criteria (i.e. $Q_{(A)B} \,|phys> = 0$) w.r.t. the (anti-)BRST invariant 
(i.e. $s_{ab} Q_{AB} = 0, \; s_b Q_B = 0 $) conserved (anti-)BRST charges $Q_{(A)B}$. First of all, we focus on the 
physicality criterion (i.e. $Q_{B} \,|phys> = 0 $) w.r.t. the BRST invariant (i.e. $s_b Q_B = 0 $) {\it modified} version of the BRST charge $Q_B$
(which has been {\it consistently} derived from the Noether BRST charge $Q_b$).
The first term (i.e. $B^a (D_0 C)^a$), as discussed earlier in the context of the physicality criterion (i.e. $Q_{b} \,|phys> = 0 $) w.r.t. the 
Noether conserved BRST charge $Q_b$, leads to the condition on the physical states where the 
operator form of the primary constraint (i.e. $\Pi^{0a}_{(A)} = B^a $)
annihilates (i.e. $\Pi^{0a}_{(A)} \, |phys> = 0 $) the physical states [cf. Eq. (51)]. On the other hand, 
the {\it second} term (i.e. $\dot B \cdot C $) consists of the physical field $\dot B^a$
(with the ghost number equal to zero) and the ghost field $C^a$ (with the ghost number equal to $+\, 1 $). Hence, the ghost field will act  on the 
ghost states (i.e. $|ghost> $) producing the non-zero result. Thus, to satisfy the physicality criterion (i.e. $Q_{B} \,|phys> = 0$), we
demand that  $\dot B^a \, |phys> = 0$. At this juncture, we invoke the validity of the
EL-EoM: $D_\mu F^{\mu\nu} - \partial^\nu B - (\partial^\nu {\bar C} \times C) = 0 $ (which is derived from the {\it perfectly} BRST-invariant 
Lagrangian density ${\cal L}_{(B)}$ w.r.t.  the gauge field $A_\mu$). If we take into account  the component $\nu = 0 $, we end-up
with $\dot B^a = - [(D_i \Pi^{i}_{(A)})^a + (\dot {\bar C} \times C)^a]$. It is clear that the second term on the r.h.s., despite carrying 
the ghost number equal to zero,  is  made-up of {\it purely} the 
(anti-)ghost fields. Hence, it will act only on the ghost states  and it will not act on the {\it physical} states 
because of obvious reasons
(which have been explained earlier\footnote{To be precise, even though the term $(\dot{\bar C} \times C)^a$ carries the ghost number equal to {\it zero}, it is 
{\it not} a physical field because it consists of (i) the anti-ghost field, and (ii) the corresponding ghost field. Hence,
this operator will {\it not} put any condition/restriction on the physical states (i.e.$|phys>$) of our BRST-quantized theory. }. Hence, ultimately, 
we obtain the following condition on the physical states 
\begin{eqnarray}\label{53}
\dot B^a \, |phys> = 0 \;\; \Longrightarrow  \;\; -\, (D_i \Pi^{i}_{(A)})^a\,  |phys> = 0 \;\;\Longleftrightarrow \;\;(D_i F^{0i})^a \, |phys> = 0,
\end{eqnarray}
where $(D_i \Pi^{i}_{(A)})^a \approx 0$ is the {\it secondary} constraint (cf. Appendix A)
on the {\it classical} version of our D-dimensional non-Abelian
1-form gauge theory. Thus, from equation (53), it is clear that the operator form of the {\it secondary} constraint annihilates the physical
states (i.e. $|phys> $) at the {\it quantum} level.  In other  words, we have observed that the BRST-invariant ($ s_b Q_B = 0 $) version of the
conserved BRST charge $Q_B$, used in the physicality criterion (i.e. $Q_{B} \,|phys> = 0 $), leads to the annihilation of the physical states
by the operator forms of the primary as well as the secondary constraints of our D-dimensional 
BRST-quantized version of the non-Abelian 1-form gauge theory (without any interaction with matter fields).

Exactly the same results will be obtained when we shall discuss the 
physicality criterion (i.e. $Q_{AB} \,|phys> = 0 $) w.r.t. the conserved and anti-BRST invariant (i.e. $s_{ab} Q_{AB} = 0 $) version of
the anti-BRST charge $Q_{AB}$. In other words, we shall obsrve the annihilation of the physical states
by the operator forms of the primary as well as the secondary constraints of our D-dimensional 
BRST-quantized version of the non-Abelian 1-form gauge theory from the requirement: $Q_{AB} \,|phys> = 0 $. However, an elaborate discussion on it 
will be {\it only} an academic exercise.  Hence, we do {\it not} devote any time on the discussion of it.\\

\vskip 0.7cm

\section{Conclusions}

In our present investigation, we have given a great deal of importance to the {\it basic} canonical (anti)commutators (associated with the BRST-quantized
version of a D-dimensional non-Abelian 1-form gauge theory) which have been utilized in the proof of (i) the Noether conserved
(anti-)BRST charges $Q_{(a)b}$ as the generators for the infinitesimal, continuous and off-shell nilpotent (anti-)BRST transformations 
(cf. Sec. 3, Appendix D), and (ii) the 
(anti-)BRST invariance (i.e. $s_{ab} Q_{AB} = 0, \; s_b Q_B = 0 $) of the (anti-)BRST charges $Q_{(A)B} $
(cf. Subsec. 5.2 for details)
of the {\it consistently} modified versions of the {\it Noether}
conserved (anti-)BRST charges  $Q_{(a)b} $ [cf. Eq. (5)]. Our present exercise corroborates the key observations of our earlier work [6] where
the ideas of (i) the infinitesimal, continuous and 
nilpotent (anti-)BRST symmetry transformations $s_{(a)b}$, and (ii) their 
generators as the  Noether conserved (anti-)BRST charges $Q_{(a)b} $, have {\it only} been 
taken into considerations. We have established the {\it superiority} of our present canonical approach over our
earlier work [6] where {\it only} the relationship (11)
 has been taken into account to prove that the {\it modified} versions of the (anti-)BRST charges $Q_{(A)B}$ are (anti-)BRST invariant. 
In other words, we have been able to prove the (anti-)BRST invariance of the conserved charges $Q_{(A)B}$ in two {\it different} ways in our present 
endeavor where (i) the symmetry considerations, and (ii) the use of the basic canonical brackets, {\it both} have been taken into account to prove:
$s_b Q_B = 0, \; s_{ab} Q_{AB} = 0$.

In our present endeavor, we have concentrated on the off-shell nilpotent (i.e. $s^2_{(a)b} = 0 $) versions of the (anti-)BRST symmetry
transformation  operators $s_{(a)b}$ and shown that the Noether conserved (anti-)BRST charges $Q_{(a)b}$ [cf. Eq. (5)] are (i) 
{\it not} the (anti-)BRST invariant quantities [cf. Eq. (12)], and (ii) {\it not} found to be off-shell nilpotent of order two [cf. Eq. (12)]. 
These conclusions have been drawn w.r.t. our focus on the relationship between the infinitesimal, continuous and off-shell nilpotent 
(anti-)BRST symmetry transformations and their generators 
as the conserved Noether (anti-)BRST charges. However, we have
established the importance of the Noether conserved (anti-)BRST charges (cf. Sec. 3, Appendix D) as the generators [cf. Eq. (11)]
for the off-shell nilpotent (anti-)BRST symmetry transformations (2). On the other hand, we have consistently derived the 
{\it modified} versions of the conserved (anti-)BRST charges $Q_{(A)B}$ [from the Noether (anti-)BRST charges $Q_{(a)b}$] which are found
to be (anti-)BRST invariant (i.e. $s_{ab} Q_{AB} = 0, \; s_b Q_B = 0 $) quantities
w.r.t. the {\it off-shell} nilpotent (anti-)BRST symmetry transformations (2). Thus, the (anti-)BRST charges $Q_{(A)B}$  are {\it physical} quantities
within the framework of BRST formalism. As far as the physicality criteria [5], within the purview of BRST approach to gauge theories are concerned, we have
established the {\it superiority} (cf. Subsec. 5.3)
of the conserved (anti-)BRST charges $Q_{(A)B}$ over the Noether conserved (anti-)BRST charges $Q_{(a)b}$ where we have shown that the
physicality criteria  (i.e. $Q_{(A)B} \,|phys> = 0 $) lead to the annihilations of the physical states (i.e. $|phys>$) by the operator form of the
first-class constraints of our {\it classical} 
D-dimensional non-Abelian 1-form gauge theory (cf. Appendix A) which are consistent with the
Dirac quantization conditions on the gauge systems (see, e.g. [11,12,21-25]). These  conditions are {\it not} obtained when we demand the
physicality criteria (i.e. $Q_{(a)b} \,|phys> = 0 $) w.r.t. the Noether conserved 
(anti-)BRST charges $Q_{(a)b}$ (cf. Subsec. 5.3) which are {\it not} (anti-)BRST invariant [cf. Eq. (12)].

As far as the nilpotent versions of the (anti-)BRST charges are concerned, we have shown that the Noether conserved (anti-)BRST charges $Q_{(a)b}$
are the (anti-)BRST invariant quantiles and they obey the nilpotency property, provided we exploit the theoretical strength of 
(i) the  Gauss divergence theorem at appropriate places, and
(ii) the EL-EoM w.r.t. the gauge field at suitable places (cf. Sec. 4 for details). We have corroborated these statements by exploiting (i) the beauty of 
the (anti-)BRST symmetry transformations and their generators as the Noether conserved 
(anti-)BRST charges  (cf. Subsec. 4.1 for details), and (ii) the theoretical strength of 
the Gauss divergence theorem and EL-EoM w.r.t. the non-Abelian gauge field (cf. Subsec. 4.2 for details).    Thus, as far
as the discussion on the BRST cohomology (see, e.g. [7]) is concerned, we can take into account the Noether conserved (anti-)BRST charges $Q_{(a)b}$.
 We, ultimately, conclude that
{\it both} types of the conserved (anti-)BRST charges (i.e. $Q_{(a)b}$ and $Q_{(A)B}$) have their (i) own identity
(i.e. $\{Q_b, \;Q_B \} = 0  $ and $\{Q_{ab}, \; Q_{AB} \} = 0$),  and 
(ii) own importance (cf. Subsec. 5.3 for details), within the framework of BRST formalism. In a very recent work [26], we have considered the BRST-quantized version of the
D-dimensional Abelian 2-form gauge theory that is {\it also} endowed with a {\it single} CF-type restriction as is the case with 
our present BRST-quantized
version of the D-dimensional non-Abelian 1-form gauge theory. We have been able to establish (in {\it this} recent work [26]) the nilpotency
property of (i) the Noether conserved (anti-)BRST charges, and (ii) the consistently {\it modified} versions of the (anti-)BRST charges,
by taking into account the {\it basic} canonical (anti)commutators. We have {\it not} yet been able to do the same for our present
BRST-quantized D-dimensional non-Abelian 1-form gauge theory. We envisage to take this project in our future endeavor where we wish to
prove the nilpotency and absolute anticommutativity properties of the conserved (anti-)BRST charges 
by taking into account the basic canonical brackets of our present BRST-quantized theory.

In our earlier work [6], we have discussed the Noether theorem and nilpotency property of the (anti-)BRST charges,
in the context of (i) the 1D SUSY toy model
of a spinning relativistic particle, and (ii) the {\it massless} and St{\" u}ckelberg-modified 
{\it massive} D-dimensional Abelian 2-form and 3-form gauge theories
(besides our present D-dimensional BRST-quantized non-Abelian 1-form gauge theory).
All these systems are endowed with the {\it non-trivial} CF-type restriction(s). As a consequence, we have found that the Noether conserved (anti-)BRST
charges are non-nilpotent  as well as non-(anti-)BRST invariant quantities
w.r.t. the {\it off-shell} nilpotent (anti-)BRST transformation operators. We have found the 
expressions for the (anti-)BRST invariant  (i.e. $s_{ab} Q_{AB} = 0, \; s_b Q_B = 0 $) versions of the (anti-)BRST charges $Q_{(A)B}$ from the 
 Noether (anti-)BRST charges $Q_{(a)b}$ by using (i) the appropriate EL-EoM, and
(ii) the Gauss divergence theorem. It would be nice future endeavor to focus on these 
{\it additional} physical systems and apply our present canonical approach to
prove (i) the nilpotency of the Noether conserved (anti-BRST charges by using the Gauss divergence theorem, and (ii) the (anti-)BRST invariance of the {\it consistently} modified versions $Q_{(A)B}$ of
the Noether (anti-)BRST charges $Q_{(a)b}$. The thorough discussion on the physicality criteria w.r.t. the {\it above} (anti-)BRST 
invariant versions of the (anti-)BRST charges  $Q_{(A)B}$ is yet another interesting direction for our future investigation(s).  \\

\vskip 0.5cm



\noindent
{\bf Data availability}\\

\vskip 0.3cm

\noindent
No data was used for the research described in this article.\\

\vskip 0.3cm

\noindent
{\bf Declaration of competing  of interest}\\

\vskip 0.3cm

\noindent
The author declares that he has no known competing financial interests or personal relationships that could have 
appeared to influence the work reported in this paper.\\






\vskip 0.7cm

\begin{center}
{\bf Appendix A: \bf On the First-Class Constraints of the Classical Theory}
\end{center}

\vskip 0.7cm

\noindent
Our {\it classical} D-dimensional non-Abelian 1-form gauge theory is endowed with a set of {\it two} first-class constraints
in the terminology of Dirac's prescription for the classification scheme of constraints [11,12,21-25]. We discuss these 
constraints, very briefly, in our present Appendix because they appear {\it explicitly} in the physicality criteria w.r..t the 
(anti-)BRST invariant versions of the conserved
(anti-)BRST charges of our theory (see, e.g. Sec. 5). Toward this goal in mind, 
we begin with the following Lagrangian density (${\cal L}_{(0)} $) of our {\it classical} D-dimensional non-Abelian 1-form gauge theory (which has the 
self-interaction due to its non-Abelian nature but has  {\it no} interaction with the matter fields), namely;
\[
{\cal L}_{(0)} = -\, \frac{1}{4}\, F^{\mu\nu} \cdot F_{\mu\nu} \equiv -\, \frac{1}{4}\, F^{\mu\nu a}  F^{a}_{\mu\nu},
\eqno(A.1)
\]
where the field strength tensor $F_{\mu\nu}^a = \partial_\mu\, A_\nu^a - \partial_\nu\, A_\mu^a + i\, f^{abc}\, A_\mu^b\, A_\nu^c$
($\mu, \nu...= 0, 1, 2...D-1 $) is for the non-Abelian 1-form $A^{(1)} = A_\mu \, d x^\mu\equiv A_\mu^a\, T^a\,d\, x^\mu $ gauge field $A_\mu^a$.
The SU(N) generators $T^a$ obey the algebra $[T^a, \; T^b ] = f^{abc}\, T^c $ (with $ a, b, c... = 1, 2, 3...N^2-1)$) where the 
structure constants $ f^{abc} $ are totally antisymmetric in all the indices (see, e.g. [7]). The canonical conjugate momenta ($\Pi^{\mu a}_{(A)}$) w.r.t.
the gauge field $A_\mu^a$ are as follows
\[
\Pi_{(A)}^{\mu a}  = \dfrac{\partial \, {\cal L}_{(0)}}{\partial\, (\partial_0\, A^a_\mu)} = -\, F^{0\mu a} 
\;\;\Longrightarrow\;\; \Pi_{(A)}^{0 a}  = -\, F^{00 a} \approx 0 , \qquad  \Pi_{(A)}^{i a}  = -\, F^{0i a}, 
\eqno(A.2)
\]
where $\Pi_{(A)}^{0 a}  \approx 0 $ is the primary constraint (PC) on our theory. The symbol $\approx $ stands for 
{\it weakly} zero in the Dirac notation  
for the discussion on the topic of constraints (see, e.g. [11,12,21-25] for details). As a consequence, we 
are allowed to take a {\it first-order} time derivative on PC
to obtain the secondary constraint (SC) on our theory. In other words, we can 
demand the time-evolution invariance (i.e. $\partial_0\, \Pi_{(A)}^{0 a}  \approx 0 $)
of the PC to obtain the secondary constraint. For such derivation, the most suitable method is the Hamiltonian approach
(see, e.g. [11,12,21-24]). However, for our simple and {\it well-studied} system, we can utilize the theoretical strength of the EL-EoM 
w.r.t. the gauge field $A_\mu^a$ and demand  the
time-evolution invariance of the PC  to obtain the SC on our theory (see, e.g. [7]). At this juncture, it is worthwhile to point out that these constraints
exist on our theory because the starting Lagrangian  density $(A.1)$ is {\it singular} (see, e.g. [12,21-25]).

Against the backdrop of the above discussion, we note that the EL-EoM [that emerges out 
from the Lagrangian density $(A.1)$] w.r.t. the gauge field $A_\mu^a$  is as follows:
\[
(D_\mu\, F^{\mu\nu})^a = 0  \;\;\Longrightarrow\;\; \partial_\mu F^{\mu\nu a} + i\, (A_\mu \times F^{\mu\nu})^a = 0.
\eqno(A.3)
\]
Taking into account the choice: $\nu = 0 $,  we obtain the following form of the above EL-EoM
\[
(D_0\, F^{00})^a + (D_i\, F^{i0})^a = 0  \;\;\Longrightarrow\;\; \partial_0 F^{00 a} + i\,(A_0 \times F^{00})^a + (D_i\, F^{i0})^a = 0,
\eqno(A.4)
\]
where the term $i\,(A_0 \times F^{00})^a $  will be zero due to $F^{00} = 0$ which is implied by the antisymmetric 
(i.e. $F^{\mu\nu} = -\, F^{\nu\mu} $) nature  of $F^{\mu\nu} $. To corroborate this assertion, we would like to add that,
according to Dirac's prescription, we are allowed to take {\it only} the first-order time derivative on the PC {\it otherwise} the 
the explicit expression for PC (without a first-order time derivative) is {\it strongly} equal to zero.
Hence, the {\it second} term, present in the above equation $(A.4)$, will be zero. In the language of the canonical conjugate 
momenta [cf. Eq. $(A.2)$], the above equation $(A.4)$ leads to the following mathematical relationship:
\[
\partial_0 \, \Pi_{(A)}^{0 a}  = (D_i\, \Pi_{(A)}^{i})^a  \equiv \partial_i \Pi_{(A)}^{ia} + i\, (A_i \times \Pi_{(A)}^{i})^a \approx 0.
\eqno(A.5)
\]
In other words, the requirement of the time-evolution invariance [i.e. $\partial_0 \, \Pi_{(A)}^{0 a} \approx 0 $] leads to the derivation of 
the secondary constraint on our theory as: $(D_i\, \Pi_{(A)}^{i})^a \approx 0 $.

For our present system (which has been studied using various
theoretical methods), it has been found that there are no further constraints on our theory. Since {\it both} the constraints [i.e. $\Pi_{(A)}^{0 a} \approx 0 $ 
and $(D_i\, \Pi_{(A)}^{i})^a \approx 0 $] are expressed in terms of the
components of the conjugate momenta [cf. Eq. $(A.2)$], they commute with each-other thereby rendering them to belong to the first-class constraints 
in the terminology of Dirac's prescription for the classification scheme of constraints (see, e.g. [11,12,21-25]). Before we end this 
Appendix, as a side remark, we would like to state that the {\it Abelian} limit of the secondary constraint [i.e. $(D_i\, \Pi_{(A)}^{i})^a \approx 0 $]
is nothing but the Gauss law of the source-free Maxwell's equations.\\

\vskip 0.7cm

\begin{center}
{\bf Appendix B: \bf On the CF-Type Condition}
\end{center}

\vskip 0.7cm

\noindent
The {\it classical} gauge theories are characterized by the existence of the first-class constraints on them 
in the terminology of Dirac's prescription for the classification scheme of constraints (cf. Appendix A for details). 
In exactly similar fashion, the existence of the (non-)trivial CF-type restriction(s)
is the hallmark\footnote{Some of the examples 
for the BRST-quantized theories (see, e.g. [27] for details)
that are endowed with the {\it trivial} CF-type
restriction are (i) the D-dimensional Abelian 1-form gauge theory, (ii) the toy model of a rigid rotor, (iii) the
free scalar relativistic particle, etc. }
of a given BRST-quantized gauge theory. We have been able to corroborate this statement in our earlier
works (see, e.g. [28,29] ) on the  BRST-quantized versions of (i) the D-dimensional non-Abelian 1-form gauge theory,
and (ii) the Abelian 2-form and 3-form gauge theories, where we have established the deep connections between
the non-trivial CF-type restrictions and the geometrical objects called gerbes.
The key objective of our present short Appendix is to show that the {\it non-trivial} CF-condition [15] exists on
our BRST-quantized D-dimensional non-Abelian 1-form gauge theory by taking into considerations (i) the direct
equality (i.e. ${\cal L}_{(B)} = {\cal L}_{(B)} $) 
of the coupled (but equivalent) Lagrangian densities [cf. Eq. (1)] by demanding that:
 ${\cal L}_{(B)} - {\cal L}_{(\bar B)} = 0 $ 
(modulo a total spacetime derivative), and (ii) the requirement that {\it both} the Lagrangian densities must respect the BRST  as well as 
the anti-BRST symmetry transformations {\it together}.

Against the backdrop of the above paragraph, first of all, we note that the direct {\it equality} \ of the Lagrangian densities 
in equation (1) implies the following 
\[
{\cal L}_{(B)} - {\cal L}_{(\bar B)} = (\partial_\mu A^\mu) \cdot [B + \bar B + (\bar C \times C)]  - \partial_\mu [A^\mu \cdot (\bar C \times C)],
\eqno(B.1)
\]
which demonstrates that the coupled Lagrangian densities ${\cal L}_{(B)}$  and ${\cal L}_{(\bar B)}$ are {\it equal} (modulo a total spacetime
derivative term) on the quantum field space where the CF-condition: $B + \bar B + (\bar C \times C) = 0 $ is satisfied. In other words, it is 
because of the presence of the 
CF-condition [15] on our BRST-quantized theory that the  coupled Lagrangian densities ${\cal L}_{(B)}$  and ${\cal L}_{(\bar B)}$ are
{\it equivalent}. As far as the symmetry considerations are concerned, we have seen that the action integrals, corresponding to the 
Lagrangian densities ${\cal L}_{(B)}$  and ${\cal L}_{(\bar B)}$, respect {\it perfect} BRST and anti-BRST symmetry transformations (2),
respectively. In addition, we observe the following transformations of the Lagrangian densities ${\cal L}_{(B)}$  and ${\cal L}_{(\bar B)}$, namely;
\[
s_b{\cal L }_{(\bar B)}\;  =  \partial_\mu\,[ {\{ B + (\bar C\times C)\}}\cdot
\partial^\mu C \,]-{\{B + \bar B + (\bar C\times C)\}}\cdot D_\mu\partial^\mu C,
\]
\[
s_{ab}{\cal L}_{(B)} = -\;\partial_\mu\,[{\{\bar B + (\bar C\times C)\} \cdot \partial^\mu \bar C}\,] 
 +  \;\{(B+\bar B + (\bar C\times C)\} \cdot D_\mu \partial^\mu \bar C,
\eqno(B.2)
\]
which demonstrate that the action integrals, corresponding to {\it both} the Lagrangian densities ${\cal L}_{(B)}$  and ${\cal L}_{(\bar B)}$,
respect {\it both} the off-shell nilpotent symmetry transformations of equation (2) on the space of quantum fields where the 
CF-condition: $B + \bar B + (\bar C \times C) = 0 $ is satisfied.

We end this short (but quite important) Appendix with the following concluding remark. It is worthwhile to point out that 
one of the sacrosanct requirements of the BRST and anti-BRST symmetry transformation operators is the validity of the 
absolute anticommutativity property (i.e. $\{ s_b, \; s_{ab} \} = 0 $). This key property is satisfied if and only if we
invoke the sanctity of the CF-condition because we observe the following:
\[
\big \{ s_b, \; \; s_{ab} \big \} \, A_\mu = i\,D_\mu\,\big [B + \bar B + (\bar C \times C) \big ], \quad
\big \{ s_b, \; \; s_{ab} \big \} \, \Omega = 0, \quad \Omega = B, \bar B, C, \bar C.
\eqno(B.3)
\]
In other words, we observe that the (anti-)BRST invariant (i.e. $s_{(a)b} [B + \bar B + (\bar C \times C)] = 0 $) CF-condition
is responsible for (i) the absolute anticommutativity property (i.e. $\{ s_b, \; s_{ab} \} = 0 $) of the off-shell nilpotent
(i.e. $s_{(a)b}^2 = 0 $) (anti-)BRST symmetry transformation operators that distinguish {\it them} from the nilpotent ${\mathcal N} = 2 $
SUSY transformation operators, and (ii) the existence of the coupled (but equivalent) Lagrangian densities ${\cal L}_{(B)}$  and ${\cal L}_{(\bar B)}$
[cf. Eq (1)] for the BRST-quantized versions of the D-dimensional non-Abelian 1-form gauge theory. \\

\vskip 0.7cm

\begin{center}
{\bf Appendix C: On the Standard Rules for Canonical Brackets}\\
\end{center}

\vskip 0.7cm

\noindent
The key feature of our present short Appendix is to collect a few {\it basic} (anti)commutator rules that have been used in the explicit computations
of some of the important  equations in Sec. 3 and Subsec. 5.2. In this context, we use
the following {\it standard} rules of the (anti)commutators 
for the composite as well as the independent operator(s) at various stages 
of our theoretical discussions. These very useful relationships amongst the bosonic and fermionic 
(composite as well as independent) operators are as follows
\[
\big  \{ F_1\, B_1, \; F_2 \big \} = F_1 \, \big  [ B_1, \; F_2 \big ] + \big  \{ F_1, \; F_2 \big \} \, B_1, 
\]
\[
\big  \{ B_1\, F_1, \; F_2 \big \} = B_1 \, \big  \{ F_1, \; F_2 \big \} - \big  [ B_1, \; F_2 \big ] \, F_1, 
\]
\[
\big  \{ F_1, \; B_2\, F_2 \big \} =  \big  [ F_1, \; B_2 \big ] \, F_2 + B_2 \, \big  \{ F_1, \; F_2 \big \}, 
\]
\[
\big  \{ F_1, \; F_2 B_2 \big \} = \big  \{ F_1, \; F_2 \big \}\, B_2 - F_2\,  \big  [ F_1, \; B_2 \big ] , 
\]
\[
\big  [ F_1\, F_2, \; F_3 \big ] = F_1 \, \big  \{ F_2, \; F_3 \big \} - \big  \{ F_1, \; F_3 \big \} \, F_2, 
\]
\[
\big  [ F_1, \; F_2\, F_3 \big ] =  \big  \{ F_1, \; F_2 \big \} \, F_3 - F_2\, \big  \{ F_1, \; F_3 \big \},
\eqno(C.1)
\]
where a couple of operators $B_1$ and $B_2$ are bosonic 
(i.e. $ B_1\, B_2 = B_2\, B_1, \; B_1^2 \neq 0, \;  B_2^2 \neq 0$)
in nature. On the other hand, a set of {\it three} operators $F_1$, $F_2$ and $F_3$ are fermionic 
(i.e. $F_1 \,F_2 = -\; F_2 \,F_1, \, F_1 \,F_3 = -\; F_3 \,F_1, 
\, F_2 \,F_3 = -\; F_3 \,F_2,, \; F_1^2 = 0, \,  F_2^2 = 0, \,  F_3^2 = 0 $) in nature.

We end this very short Appendix with the remark that even though we have listed the above (anti)commutators for {\it only} two bosonic field
operators (i.e. $B_1, \, B_2 $) and three fermionic field operators (i.e. $F_1, \, F_2, \, F_3 $), the algebraic relationships
$(C.1)$ can be used for the computations of the (anti)commutators for multiple bosonic and fermionic field operators, too. A close and careful look at
equation $(C.1)$ shows that we have taken only {\it three} operators on the l.h.s. However, these algebraic relationships of
the above equation $(C.1)$ can be exploited when we shall be dealing with a set of $N$ number (i.e. $N > 3 $)  of operators by
recasting them in the form of $(C.1)$ with {\it judicious}
 choices of the composite operators in a specific manner. In other words, 
we can always assume a set of operators as {\it one} operator to recast any arbitrary (anti)commutator of the $N$-number 
(i.e. $N > 3 $) of field operators in the form that is similar to $(C.1)$. This process can be repeated depending on the number of
operators (i.e. $N > 3 $) to {\it finally} obtain the given (anti)commutator {\it exactly} in the form of  $(C.1)$ at the end
of all the possible assumptions/choices of the field operators.\\

\vskip 0.7cm

\begin{center}
{\bf Appendix D: On Noether (Anti-)BRST Charges as Generators}\\
\end{center}

\vskip 0.7cm

\noindent
The central aim of our present Appendix is to demonstrate {\it concisely} that the Noether conserved (anti-)BRST charges [cf. Eq. (29)]
(which are exactly the {\it same} as the expressions for $Q_{(a)b}$ in equation (5) in a different form) are the {\it true} generators for the 
infinitesimal, continuous and off-shell nilpotent
(anti-)BRST symmetry transformations [cf. Eq. (2)] for the basic {\it dynamical} fields (i.e. $A_\mu, \, C, \, \bar C $)
of our BRST-quantized theory that is described by the coupled Lagrangian densities (1). Toward this goal in mind, let us
exploit the relationship in equation (11) to show that: $s_{ab} A_\mu = D_\mu \bar C $ (i.e. $s_{ab} A_0 = D_0 \bar C,\;
s_{ab} A_i = D_i \bar C $). Using the relationship (11), we note that, we have the folowing {\it non-zero} result, namely;
\[
s_{ab} A^a_0 \,(\vec{x}, t) = - \, i\, \Big [ A^a_0 (\vec{x}, t), \; Q_{ab} \Big ] 
\]
\[
\equiv -\, i\, \int d^{D-1} y \; \Big [ A^a_0 (\vec{x}, t), \;  - \, i\, \Pi^{0b}_{(A)} (\vec{y}, t) \, \Pi^b_{(C)} (\vec{y}, t) \Big ],
\eqno(D.1)
\]
where we have not taken the parenthesis $(\bar B)$ in front of momenta $\Pi's$ for the sake of brevity.
Using (i) the basic canonical commutator: $[A^a_0 (\vec{x}, t), \;\,  \Pi^{0b}_{(A)} (\vec{y}, t)] = i\, \delta^{ab}\, \delta^{D-1} (\vec{x} - \vec{y})$,
(ii) the property of the Dirac $\delta$-function, (iii) the appropriate commutator relationship with the composite field operators,
and (iv) the volume integration over the $y$ variable, we obtain the following
expression for the anti-BRST transformation on $A^a_0$, namely;
\[
s_{ab} A^a_0 (\vec{x}, t) = -\, i\, \Pi^a_{(C)} (\vec{x}, t) \equiv (D_0 \bar C)^a 
\;\;\; \Longrightarrow \;\;\; s_{ab} A_0 = D_0 \bar C,
\eqno(D.2)
\]
where we have taken into account the expression for the canonical conjugate momenta 
$\Pi^a_{(C)} (\bar B) $ from equation (7). Exactly in the {\it above}
fashion of computation, we obtain the following expression for the anti-BRST transformation on $A^a_i$, namely;
\[
s_{ab} A^a_i\, (\vec{x}, t) = - \, i\, \Big [ A^a_i (\vec{x}, t), \; Q_{ab} \Big ]  \equiv (D_i \bar C)^a
\;\;\; \Longrightarrow\;\;\; s_{ab} A_i = D_i \bar C,
\eqno(D.3)
\]
where the {\it first} term [i.e. $\Pi^i_{(A)} (\bar B) \cdot D_i \bar C $] 
of the expression for $Q_{ab}$ [cf. Eq. (29)] contributes to the above result. It goes without saying that
we have used (i) the basic canonical commutator: $[A^a_i (\vec{x}, t), \;  \Pi^{0jb}_{(A)} (\vec{y}, t)] 
= i\, \delta^{ab}\, \delta_i^j\, \delta^{D-1} (\vec{x} - \vec{y})$,
(ii) the property of the Dirac $\delta$-function, (iii) the standard rules of the commutators where the composite field operators are present 
(cf. Appendix C), and (iv) the volume integration over the $y$ variable. From equations $(D.2)$ 
and $(D.3)$, it is crystal clear that we have obtained the {\it correct} result for the anti-BRST symmetry transformation on the {\it bosonic} 
gauge field $A_\mu$ as:  $s_{ab} A_\mu = D_\mu \bar C $.

For the sake of completeness, let us take an example of the anti-BRST symmetry transformation: $s_{ab} C = i\, \bar B $
 on the {\it fermionic} ghost field $C$. A close look at the rules [that are connected with the basic (anti)commutators
in equation (8)] show that the following expression, in terms of the integral, exists, namely;  
\[
s_{ab} C^a (\vec{x}, t) = -\, i\, \Big \{ C^a (\vec{x}, t), \; Q_{ab} \Big \} \equiv
-\, i\, \int d^{D-1} y \; \Big \{ C^a (\vec{x}, t), \;  - \, i\, \Pi^{0b}_{(A)} (\vec{y}, t) \, \Pi^b_{(C)} (\vec{y}, t) \Big \}.
\eqno(D.4)
\]
Using the proper rules of the suitable anticommutator from our Appendix C, we note that we have the following simpler and useful 
form of the above equation, namely;
\[
s_{ab} C^a (\vec{x}, t) = -\,
\int d^{D-1} y \; \Pi^{0b}_{(A)} (\vec{y}, t) \;\Big \{ C^a (\vec{x}, t), \;  \Pi^b_{(C)} (\vec{y}, t) \Big \}.
\eqno(D.5)
\]
At this crucial stage, we use (i) the basic canonical anticommutator: $\{ C^a (\vec{x}, t), \;  \Pi^{b}_{(C)} (\vec{y}, t)] 
= i\, \delta^{ab}\, \delta^{D-1} (\vec{x} - \vec{y})$,  
(ii) the property of the Dirac $\delta$-function, (iii) the appropriate rules of the anticommutator that contain the composite 
field operators (cf. Appendix C), and (iv) the volume integration over the $y$ variable, to obtain the following
\[
s_{ab} C^a (\vec{x}, t) = -\, i\, \Pi^0_{(A)} (\vec{x}, t) = i\, \bar B^a \;\;\; \Longrightarrow \;\;\; s_{ab} C = i\, \bar B,
\eqno(D.6)
\]
which proves that the Noether conserved anti-BRST charge $Q_{ab}$ [cf. Eq. (29)] are the generator for the 
 anti-BRST symmetry transformation: $s_{ab} C = i\, \bar B $. Following exactly the same kind of theoretical tricks, it is straightforward
 to check that $s_{ab} \bar C = -\, \frac{i}{2} \, (\bar C \times \bar C) $ is {\it also} true. Thus, we have derived the
 anti-BRST symmetry transformations for the basic {\it dynamical} fields: $A_\mu, \; C$ and  $\bar C $ from the relationship (11).

Against the backdrop of the derivations of the above anti-BRST symmetry transformations, we wish to end this Appendix with the remark that
similar kinds of theoretical tricks and exercises lead to the derivations of the {\it correct} BRST symmetry transformations that have been
listed in equation (2). Without going into the details of the computations, we list here {\it all} the BRST
transformations  that are connected with the basic dynamical fields: $A_\mu, \; C, \; \bar C $ by exploiting the mathematically
beautiful relationship (11), namely;
\[
s_{b} A^a_0 (\vec{x}, t) = -\, i\, \Big [ A^a_0 (\vec{x}, t), \; Q_{b} \Big ]  \equiv i\, \Pi^a_{(\bar C)} (\vec{x}, t) \equiv (D_0 \bar C)^a 
\;\;\; \Longrightarrow \;\;\; s_{b} A_0 = D_0  C,
\]
\[
s_{b} A^a_i\, (\vec{x}, t) = - \, i\, \Big [ A^a_i (\vec{x}, t), \; \;Q_{b} \Big ]  \equiv (D_i  C)^a
\;\;\; \;\;\Longrightarrow\;\;\; \;\;s_{b} A_i = D_i C,
\]
\[
s_{b} C^a (\vec{x}, t) = -\, i\, \Big \{ C^a (\vec{x}, t), \; Q_{b} \Big \} \equiv -\, \dfrac{i}{2}\, \big (C \times C \big)^a
\;\;\; \Longrightarrow\;\;\; s_{b} C = -\, \dfrac{i}{2}\, \big (C \times C \big),
\]
\[
s_{b} \bar C^a (\vec{x}, t) = -\, i\, \Big \{ \bar C^a (\vec{x}, t), \; Q_{b} \Big \} \equiv i\, \Pi^{0a}_{(A)} = i\, B^a
\;\;\; \Longrightarrow\;\;\; s_{b} \bar C = i \, B,
\eqno(D.7)
\]
which proves that the Noether BRST charge $Q_b$ is the generator 
for the off-shell nilpotent BRST symmetry transformations:  $s_b A_\mu = D_\mu C, \; s_b \bar C = i\, B, \; s_b C = - \frac{i}{2}\, \big (C \times C \big)$
 that have been listed in equation (2) for the basic dynamical fields $A_\mu, \, \bar C $ and $C$. In other words, we have shown that the Noether
 conserved (anti-)BRST charges $Q_{(a)b}$ (even though {\it not} invariant under the off-shell nilpotent symmetry transformations) are 
 useful in the sense that they are the generators for the off-shell nilpotent (anti-)BRST symmetry transformations for the basic dynamical fields
 $A_\mu, \, \bar C$ and $C$ of the coupled Lagrangian densities (1).\\

\end{document}